\documentclass{aa} 
\bibpunct{(}{)}{;}{a}{}{,} % to follow the A&A style
\usepackage[utf8]{inputenc}
\usepackage[T1]{fontenc}
\usepackage{textcomp}
\usepackage{graphicx}
\usepackage{multirow}
\usepackage{pifont}
\usepackage{makecell} % for more vertical space in cells
\setcellgapes{5pt}
\usepackage{txfonts}
\usepackage{rotating}
\usepackage[dvipsnames]{xcolor}
\usepackage{subcaption}
\usepackage{comment}
\usepackage{amsmath,bm}
\usepackage{booktabs}
\usepackage{tabularray}
\usepackage{nicematrix}
\usepackage{sepfootnotes}

\usepackage{hyperref}

\def\herschel{{\emph{Herschel\,}}}

\begin{document}

   \title{Supervised machine learning on Galactic filaments}

   \subtitle{Revealing the filamentary structure of the Galactic interstellar medium}

   \author{A. Zavagno,
          \inst{1,2}
          \and F.-X. Dup\'e\inst{3}
          \and S. Bensaid\inst{1,3} 
          \and E. Schisano \inst{4} 
          \and G. Li Causi \inst{4} 
          \and M. Gray \inst{1} 
          \and S. Molinari \inst{4} 
          \and D. Elia \inst{4}
          \and J.-C. Lambert \inst{1}
          \and M. Brescia \inst{5}
          \and D. Arzoumanian \inst{1,6}
          \and D. Russeil \inst{1}
          \and G. Riccio \inst{5}
          \and S. Cavuoti \inst{5}
          \fnmsep\thanks{F.-X. Dup\'e and S. Bensaid contributed equally to the work presented in this article.}
          }
   \institute{Aix Marseille Univ, CNRS, CNES, LAM, Marseille, France 
            %  \email{}
  \and       Institut Universitaire de France, Paris, France 
  \and Aix Marseille Univ, CNRS, LIS, Marseille, France
  \and INAF-IAPS, via del Fosso del Cavaliere 100, I-00133 Roma, Italy 
  \and INAF – Astronomical Observatory of Capodimonte, via Moiariello 16, I-80131, Napoli, Italy 
  \and Division of Science, National Astronomical Observatory of Japan, 2-21-1 Osawa, Mitaka, Tokyo 181-8588, Japan
             %\thanks{}
             }

   \date{Received May 25 2022; accepted November 20 2022}
% \abstract{}{}{}{}{} 
% 5 {} token are mandatory
 
 \abstract
   {Filaments are ubiquitous in the Galaxy, and they host star formation. Detecting them in a reliable way is therefore key towards our understanding of the star formation process.}
   {We explore whether supervised machine learning can identify filamentary structures on the whole Galactic plane. }
   {We used two versions of UNet-based networks for image segmentation. We used H$_2$ column density images of the Galactic plane obtained with \textit{Herschel} Hi-GAL data as input data. We trained the UNet-based networks with skeletons (spine plus branches) of filaments that were extracted from these images, together with background and missing data masks that we produced. We tested eight training scenarios to determine the best scenario for our  astrophysical purpose of classifying pixels as filaments.}
   {The training of the UNets allows us to create a new image of the Galactic plane by segmentation in which pixels belonging to filamentary structures are identified. With this new method, we classify more pixels (more by a factor of 2 to 7, depending on the classification threshold used) as belonging to filaments than the spine plus branches structures we used as input. New structures are revealed, which are mainly low-contrast filaments that were not detected before. We use standard metrics to evaluate the performances of the different training scenarios. This allows us to demonstrate the robustness of the method and to determine an optimal threshold value that maximizes the recovery of the input labelled pixel classification. }
   {This proof-of-concept study shows that supervised machine learning can reveal filamentary structures that are present throughout the Galactic plane. The detection of these structures, including low-density and low-contrast structures that have never been seen before, offers important perspectives for the study of these filaments.}

   \maketitle
%-------------------------------------------------------------------
\section{Introduction}
The \textit{Herschel} infrared Galactic Plane Survey, Hi-GAL \citep{Mol10b}, revealed that the cold and warm interstellar medium (ISM) is organized in a network of filaments in which star formation is generally observed above a density threshold corresponding to $A_{\rm{V}}$=7~mag \citep{and14,kon20}. The most massive stars are formed at the junction of the densest filaments, called hubs \citep{kum20}. Because filaments host star formation and link the organization of the interstellar matter to the future star formation,  studying them is central to our understanding of all properties related to star formation, such as the initial mass function, the star formation rate, and the star formation efficiency. Filaments are therefore extensively studied with observations at all wavelengths and numerical simulations \citep[and references therein]{And10,Mol10b,Arz11,Hac18,Arz19,Shi19,Cla20,Pri22,Hac22}. All these data reveal the complex structure of filaments and show a changing morphology, depending on the way (resolution or tracers) in which they are observed \citep{leu19}. For example, high-resolution molecular line observations of Galactic filaments with the Atacama Large Millimeter Array (ALMA) show that they are made of fibers on a spatial scale < 0.1 pc \citep{Shi19,Hac18}. Their complex morphology and dynamics are  also revealed with 3D spectroscopic information \citep{mat18,hac20} and show their key role in the accretion process from large (> 10 pc) to subparsec scales, funnelling material down to the star-forming cores.  
However, the way filaments form and evolve in the ISM is still debated \citep{hoe21,hsi21}. Recent results suggest that compression from neutral (H\,{\sc{i}}) and ionized (H\,{\sc{ii}}) shells could play an important role in forming and impacting the evolution of Galactic filaments \citep{zav20,bra20}. Their detection in nearby galaxies, where they are also clearly linked to the star formation process, makes studying filaments even more important and universal \citep{fuk19}. \\
Different algorithms are used to extract filaments from 2D images \citep{sou11,sch14,koc15,zuc18,schisano2020,men21} and from 3D spectral data cubes \citep{sou11,che20}. These algorithms often rely on a threshold definition (for intensity or column density). Nonetheless, a close visual inspection of 2D images and 3D cubes show that some filaments are missed by all these detection algorithms, especially when the filaments have low column density contrasts. This means that even large surveys of the Galactic plane cannot deliver a complete (unbiased) view of the filaments present there. Another limitation of these algorithms comes from their computation time, which can make some of them too expensive to envision a complete run on large-scale surveys data for multiple defined threshold and extraction parameters. Because the multi-wavelength information available on our Galaxy on all spatial scales is so rich, proposing another way of extracting filaments might allow a leap forward for an unbiased census of these data. In this paper, we explore the potential of supervised machine learning as a new way to reveal filaments from 2D images. Using Hi-GAL data, \citet{schisano2020} extracted filaments from column density (N$_{\rm{H_2}}$) images of the Galactic plane. Based on these data, we study the possibility for convolutional UNet-based networks \citep{fu1981survey} to identify pixels as belonging to the filament class, based on the input information given as previously identified filament masks from \citet{schisano2020}. Except for the catalog of filament candidates published in \citet{schisano2020} where faint filaments present in the Galactic plane are known to be missed, no complete extraction of filamentary structures in the Galactic plane exists so far. This fact motivates our work, in which we propose an alternative method that could allow us to go beyond the current possibilities. However, this fact also indicates that we worked with an incomplete ground truth (see Section~\ref{datasetbuilding}) that renders an absolute evaluation of the method performances proposed here impossible. Nonetheless, we show that new filaments revealed by the UNet-based algorithm and not detected before are confirmed through imaging at other wavelengths, which gives us confidence that this new method can progress toward an unbiased detection of filaments.\\    
The paper is organized as follows: in Section~\ref{data} we describe the images and the information on the filament locations we used in the supervised learning. The supervised learning itself is described in Section~\ref{ml}. Results are presented in Section~\ref{res} and are discussed in Section~\ref{dis}. Conclusions are given in Section~\ref{conc}.
%---------------------------------------------------------
\section{Data}\label{data}
\subsection{Hi-GAL catalog} \label{HiGAL}
\begin{figure*}[tb]
   %\centering
\includegraphics[width=\linewidth]{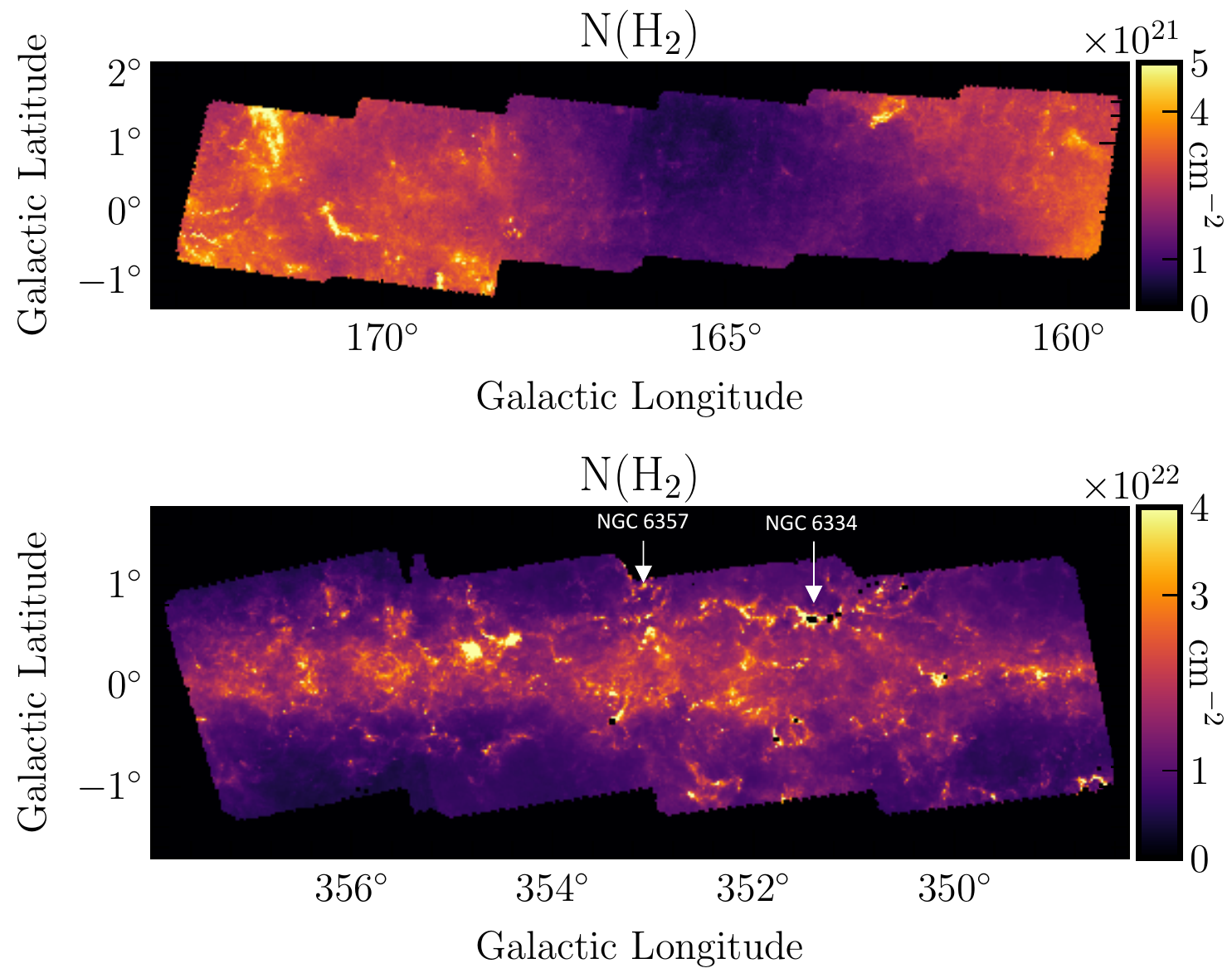}
   \caption{Hi-GAL H$_2$ column density image of the $l$=160-171\textdegree{} (top) and $l$=349-356\textdegree{} (bottom) regions produced using Hi-GAL images as described in \citet{eli13}. The $l$=349-356\textdegree{} zone contains the bright star-forming regions NGC~6334 and NGC~6357. These two regions are used in this paper to illustrate the results. The dark pixels are saturated. }
              \label{nh2}%
\end{figure*}
\begin{figure}[htb]
   \centering
\includegraphics[scale=0.6]{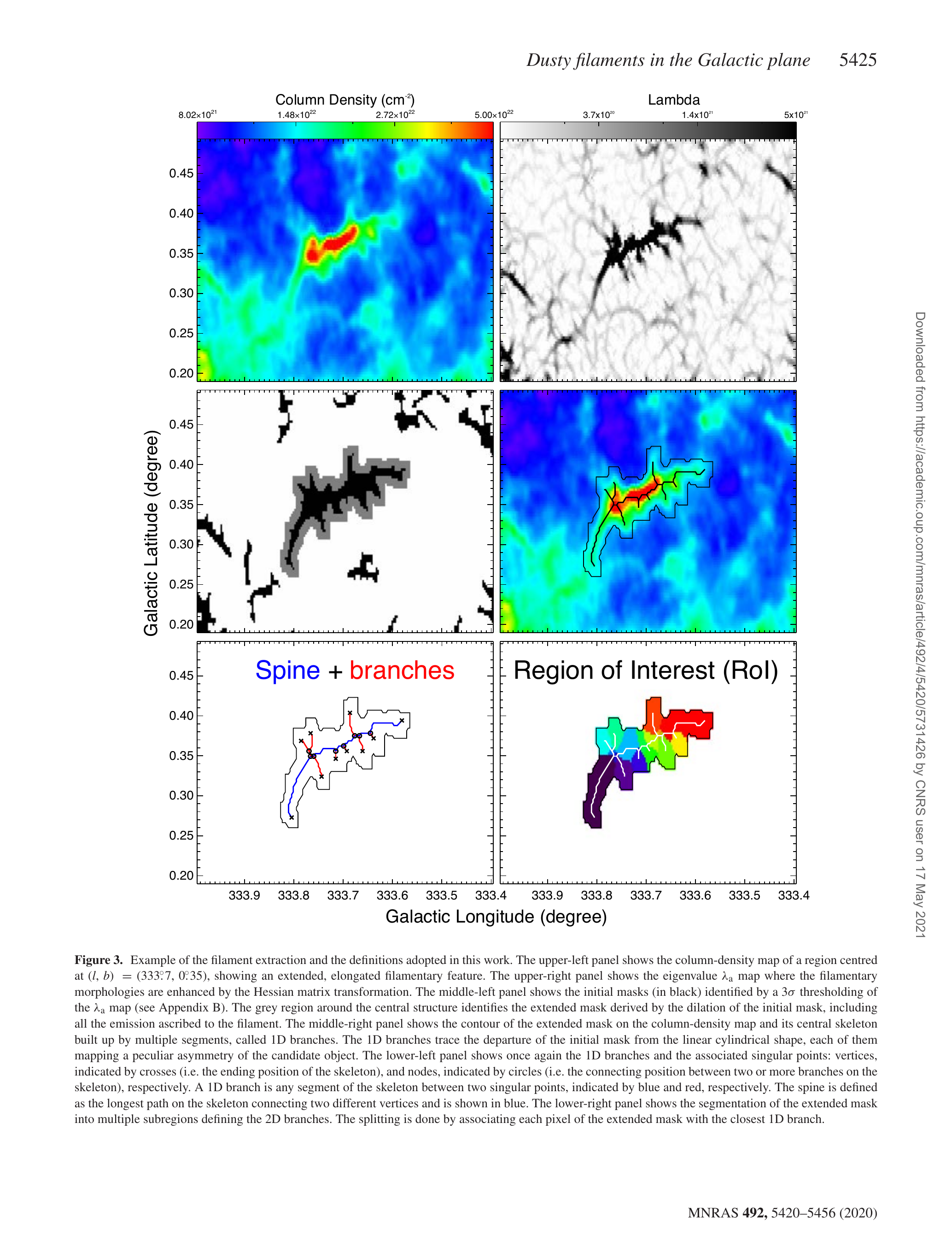}
   \caption{Illustration of the filament extraction method from \citet[][their Figure~3]{schisano2020}. The spine (blue line) and branches (red lines) associated with a filament used for the supervised training process are shown in the bottom left corner. The RoI (bottom right corner) is the zone used by \citet{schisano2020} to define a filament. }
              \label{fil}%
    \end{figure}

The \herschel infrared Galactic Plane Survey, Hi-GAL \citep{Mol10b}, is a complete survey of the Galactic plane performed in five infrared photometric bands centered at 70, 160, 250, 350, and 500\,$\mu$m. H$_2$ column density (N$_{\rm{H_2}}$) images were created for the whole Galactic plane following the method described in \citet{eli13} and \citet{schisano2020}. N$_{\rm{H_2}}$ and dust temperature maps were computed from photometrically calibrated images. The \herschel data were convolved to the 500\,$\mu$m resolution (36\arcsec), and a pixel-by-pixel fitting by a single-temperature graybody was performed. An example of the column density image covering the $l$=349-356\textdegree{} region is presented in Figure~\ref{nh2}. This region contains the bright and well-studied Galactic star-forming regions NGC~6334 and NGC~6357. \\
\citet{schisano2020} analyzed the whole Galactic plane by extracting filamentary structures from the H$_2$ column density (N$_{\rm{H_2}}$) maps.  In their work, a filament is defined as a two-dimensional, cylindric-like structure that is elongated and shows a higher brightness contrast with respect to its surroundings. The extraction algorithm is based on the Hessian matrix $H(x,y)$ of the intensity map N$_{\rm{H_2}}(x,y)$ to enhance elongated regions with respect to any other emission. The algorithm performs a spatial filtering and amplifies the contrast of small-scale structures in which the emission changes rapidly. Further filtering allows identifying the filamentary structures. 
Figure \ref{fil} shows an example of this filament extraction, reproducing the figure 3 of \citet{schisano2020}. We chose this figure because it shows the input we use in this work: the spine (blue line) and the branches (red lines, both shown in the bottom left panel) associated with a given filament.  \citet{schisano2020} defined a filament as traced by its associated region of interest (RoI; bottom right), which covers a larger area than the region that is defined with the spine plus branches. In this work we use this spine plus branches structure to define a filament because the early tests we made to train the networks with the input RoIs returned filamentary structures that were too large compared to the structure that is observed in the column density mosaics. This point is illustrated in Figure~\ref{Unet-RoI-spines} and is discussed in Sect.~\ref{sec:AstroResults}. \\
The analysis of the extracted structures from \citet{schisano2020} resulted in the publication of a first catalog of 32 059 filaments that were identified over the entire Galactic plane. We used this published catalog of filaments and their associated spine plus branches as ground truth of the filament class for the training process (see Sect. \ref{datasetbuilding}). The method is described in Section~\ref{ml}.  

\subsection{Data preprocessing} \label{datasetbuilding}
\begin{figure*}
    %\centering
    \includegraphics[width=\linewidth]{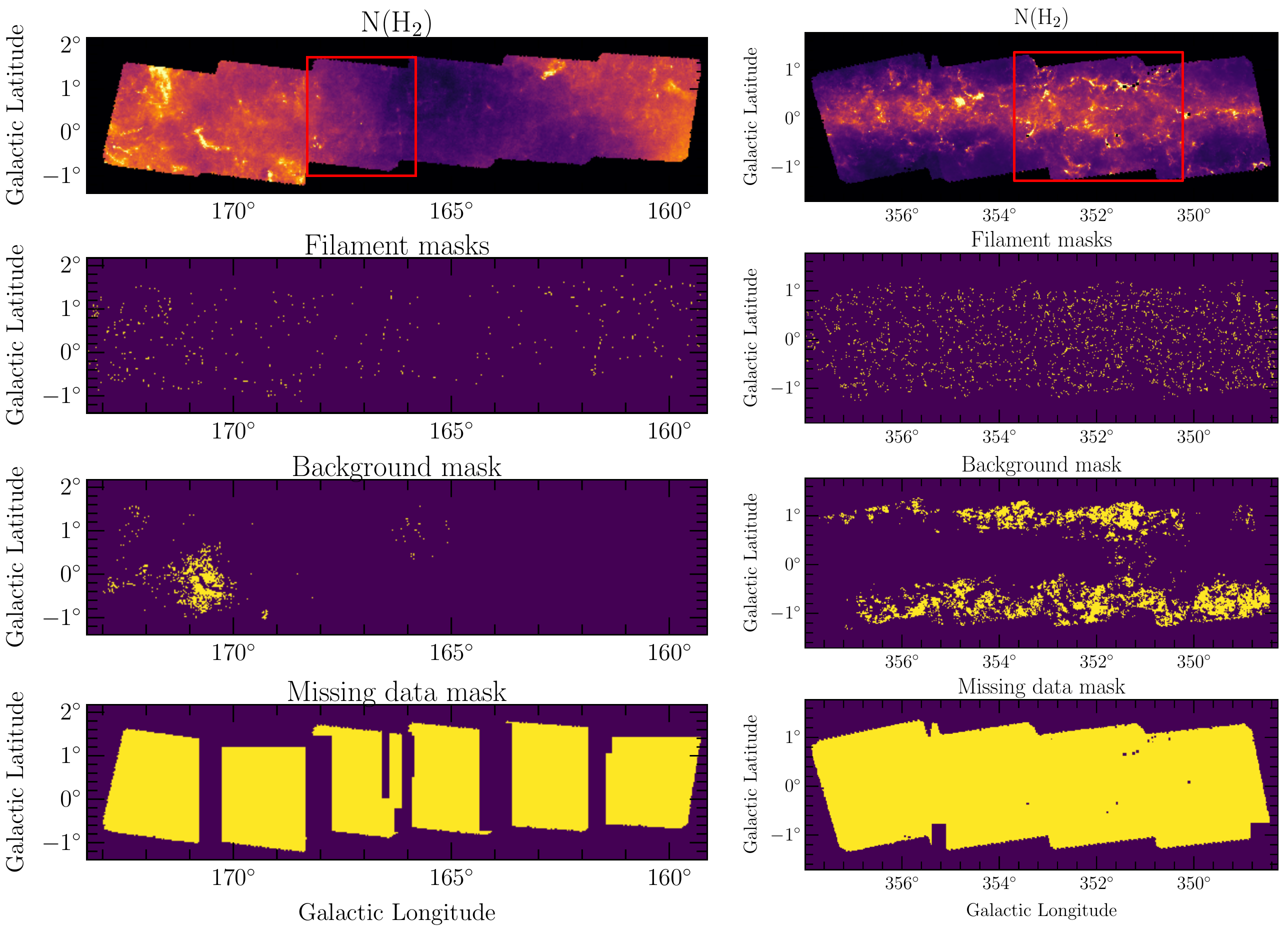}
    \caption{Illustration of the Galactic regions located at 160-171\textdegree{} (left) and 349-356\textdegree{} (right) of the four input maps used for the supervised learning. From top to bottom, we show the N$_{\rm{H_2}}$ column density map, the input filament masks, a background localization map, and the missing-data map (0 in purple, 1 in yellow). All these maps were obtained as explained in Appendix~\ref{appendixA}. The filament and background mask maps are multiplied by the missing-data map before they were used in the training process. The red rectangle shown in each column density map represents the region we extracted to compute the performance of the training (see Sect.~\ref{sec:ScoresAnalysis}).}
    \label{methodomaps}
\end{figure*}

As methods based on deep learning strongly depend on the nature and on the representativity of the input data, we took particular attention to the construction of the data set. We used four input maps (see Fig.~\ref{methodomaps}): 1) the N$_{\rm{H_2}}$ mosaics obtained as part of the Hi-GAL survey products \citep[e.g., Fig.~\ref{nh2}]{mol16,schisano2020}, 2) the spines plus branches of the detected filaments from  \citet[e.g., Fig.~\ref{fil} bottom left corner]{schisano2020}, 3) a background map (localization of nonfilament pixels), and 4) a missing-data map (see Appendix~\ref{appendixA}).
The origin and the ways in which these maps were obtained are presented in Appendix~\ref{appendixA}.  
An example of these maps is given on Figure~\ref{methodomaps}, illustrated for the two portions of the Galactic plane that are located at 160-171\textdegree\ and 349-356\textdegree. We obtained results for the whole Galactic plane, which we illustrate in two regions that we selected because they represent the diversity of column density and filaments content observed in the Galactic plane well. The 160-171\textdegree\ region samples a low column density medium (up to 8$\times$10$^{21}$ cm$^{-2}$) in which only a few filaments are detected, while the 349-356\textdegree\ region shows a rich content in filaments that are detected in a high column density medium (up to 9$\times$10$^{22}$ cm$^{-2}$, see Figure~\ref{methodomaps}).  
For the training step, we merged all the individual mosaics (10\textdegree-long in longitude direction) into one global map using the \textit{reproject} module for \textit{astropy}~\citep[][and Appendix~\ref{appendixA}]{Robitaille2020reproject}. 

\begin{figure}[htp]
    \includegraphics[width=0.8\linewidth]{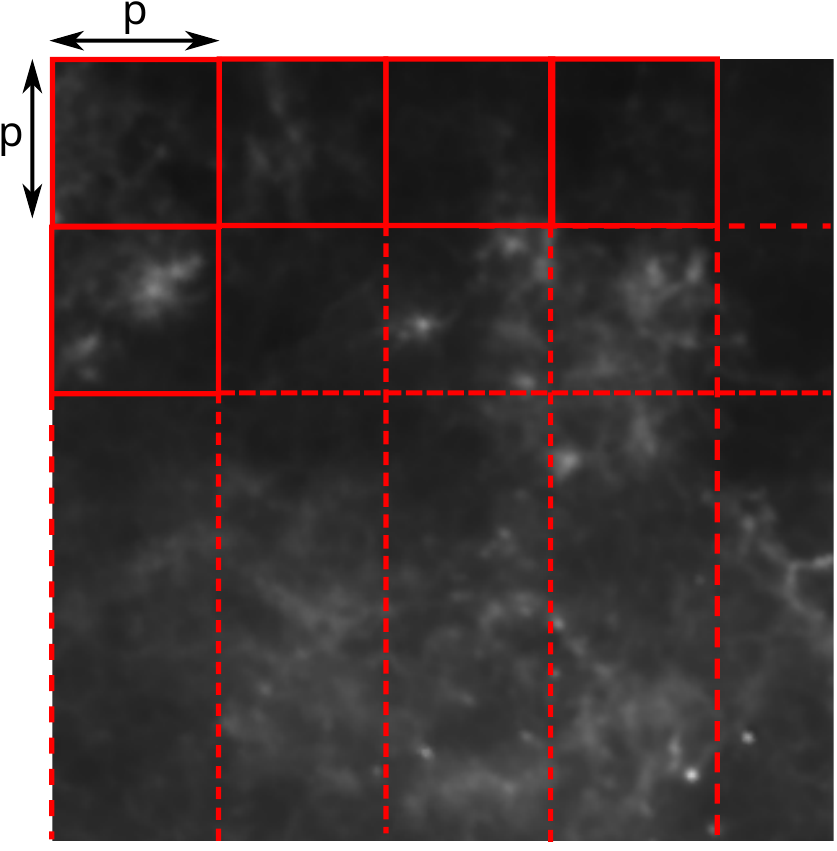}
    \caption{Construction of patches of size $p\times p$ using a sliding window.}
    \label{fig:patches_overlap}
\end{figure}
\begin{figure}
    \centering
    \includegraphics[width=0.9\linewidth]{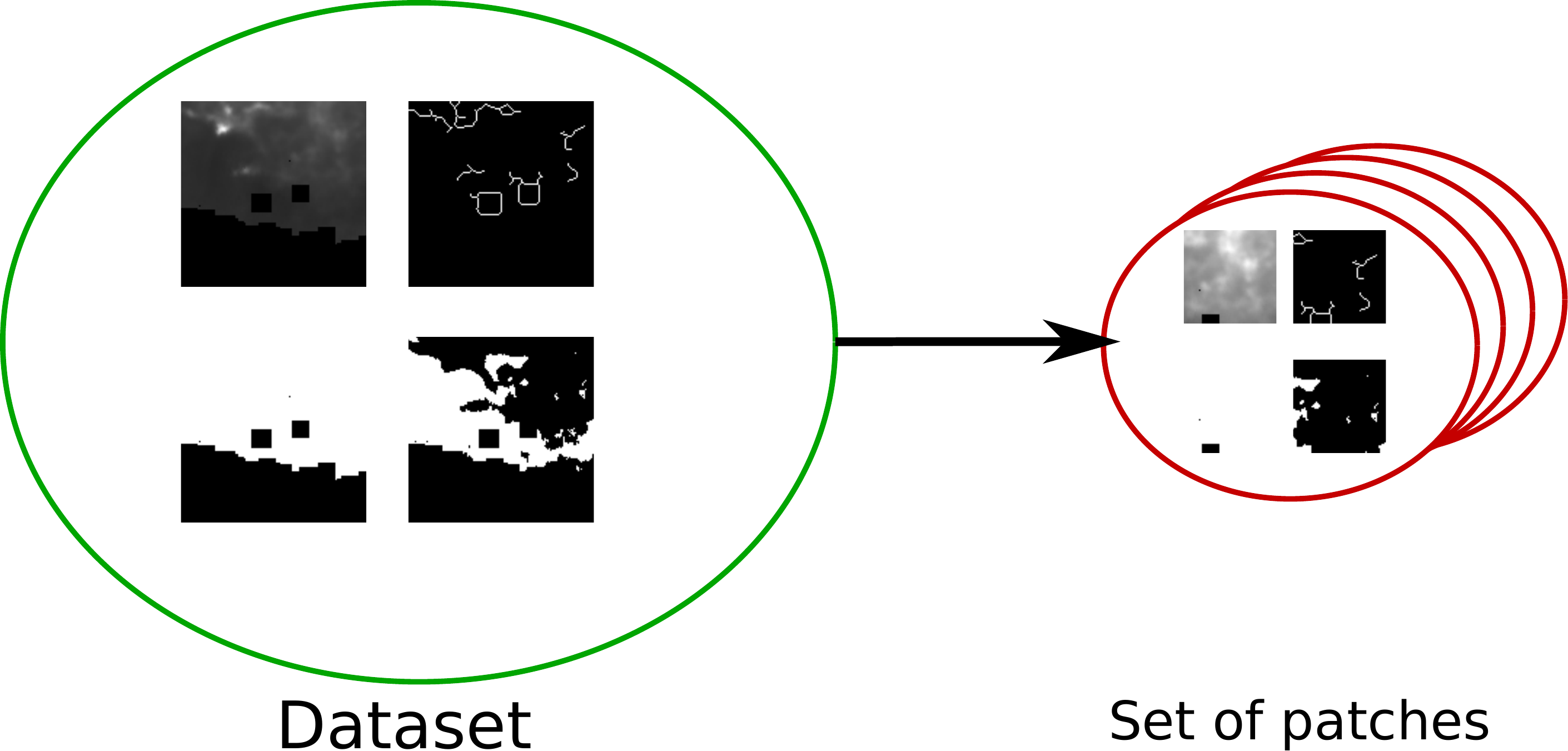}
    \caption{Building the data set using the four input maps (on the left) into a set of patches (on the right). On the left, the maps are the column density (top left), filament spine+branches (top right), missing data (bottom left), and background pixels (bottom right).}
    \label{fig:dataset}
\end{figure}

As the four input maps are very large ($150000$ $\times$ $2000$ pixels), we split them into many patches that constitute the original data set. As shown in Figure~\ref{fig:patches_overlap}, we split the maps into $p\times p$ patches. The size of $p=32$ pixels was chosen to preserve the information on the small filamentary structures. This size is also the minimum size accepted by the UNet architecture. The patches were generated by applying a sliding window of size $p$ (the patch size) to the global mosaic of H$_2$ column density (N$_{\rm{H_2}}$). To ensure the coherence between the four input maps, the four patches (N$_{\rm{H_2}}$, spine+branches, background, and missing data) were taken using the same coordinates (see Figure~\ref{fig:dataset}). In order to avoid any common information between patches, the construction was made without any overlap between the patches.
\subsection{Data augmentation}
\begin{figure}[htb]
    \includegraphics[width=0.9\linewidth]{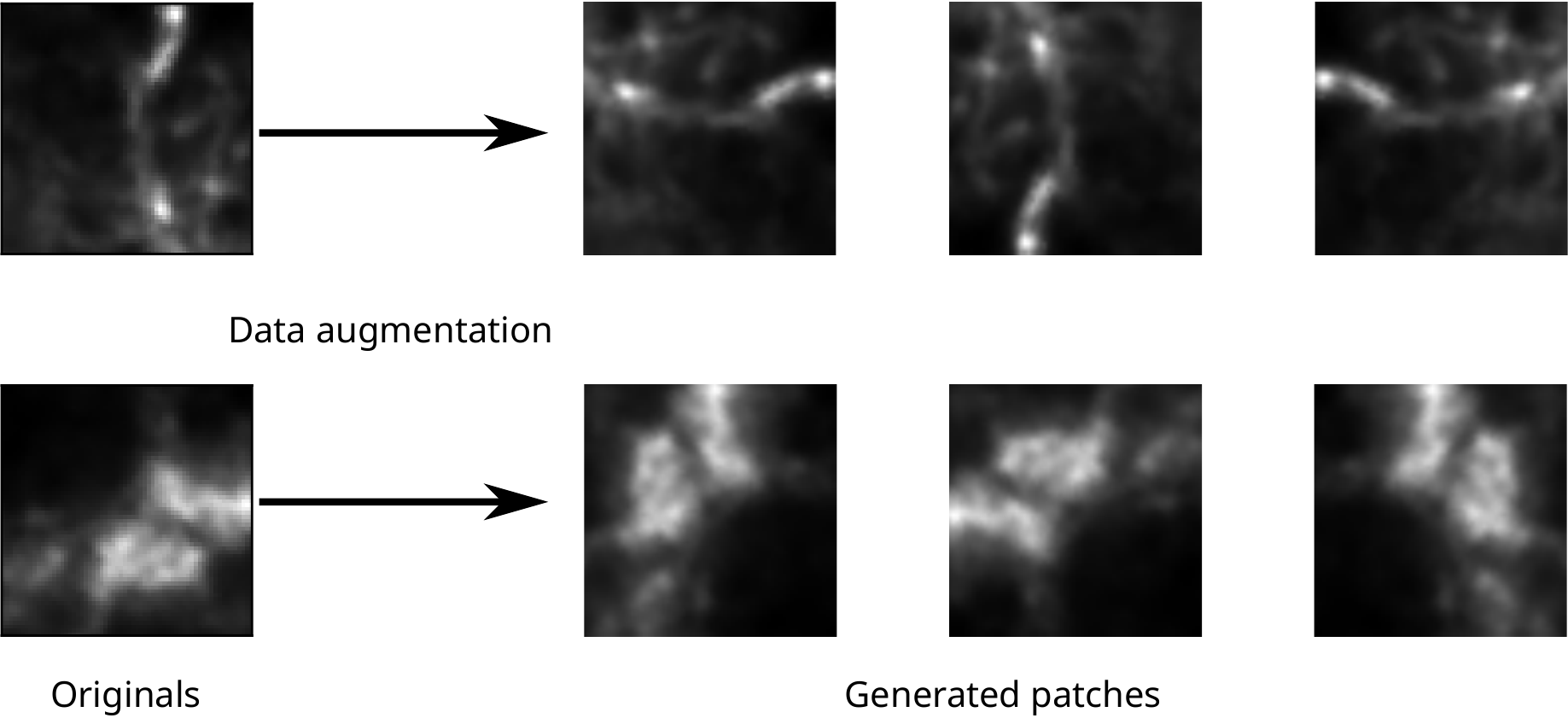}
    \caption{Example of data augmentation results.}
    \label{fig:data_augmentation}
\end{figure}

Deep neural networks are greedy algorithms. In spite of the huge size of the Galactic map, the final data set had merely $52.000$ patches after empty patches were removed. Here, we refer as “empty patches” to patches that only contain missed values (patches located on edges) or to patches that contain “0 pixels labeled data”. Fully unlabeled patches were removed from the patch data set. As unlabeled pixels represent more than $80\%$ of the data set, this resulted in the loss of many patches. The data augmentation is thus necessary in order to increase the number of training and validation patches and to thereby enable a sufficient convergence of the neural network during the training step~\citep{goodfellow2016deep}.
Two types of rotation were used: rotation around the central pixel of the squared  patch (0\textdegree (original), 90\textdegree, 180\textdegree{} , and 270\textdegree), and flipping the patch with respect to the $x$ - and $y$ -axes. We also allowed a composition of both rotations. All possible transformations are equally probable, that is, we selected the applied transformation following a uniform distribution.
To attenuate redundancy issues, the augmentation was done \textit{\textup{on the fly,}} meaning that at each batch, we produced a new set of patches using the augmentation process. With our setting, we virtually increased the number of patches by a factor equal to 64. Figure~\ref{fig:data_augmentation} shows some examples of the data augmentation process.

\section{Method}\label{ml}
\subsection{Segmentation pipeline}
Our segmentation method relies on three components: a data preparation procedure, a neural network with an architecture dedicated to the recognition of filamentary structures, and a training procedure adapted to the N$_{\rm{H_2}}$ data. After the neural network was trained, we used it to segment the N$_{\rm{H_2}}$ map. The result of the segmentation process is a map in which the pixels are classified into two classes: either a filament pixel (identified as class 1), or a background pixel (identified as class 0). With these two classes, we produced an intensity map with values of 0 and 1. The values of the classification indicate whether a pixel belongs to the filament class (the reverse map shows the classification value according to which a pixel belongs to the background class).

\subsubsection{Segmentation with UNets}
Automatic segmentation is a well-known issue in  the artificial intelligence community. Its origins lie in computer vision. It is a well-studied problem today, especially where segmentation is mandatory for decision or prediction, typically for medical or biology images~\citep{fu1981survey, alzahrani2021biomedical}. It has also been used for a long time in astrophysics, with recent applications on galaxies \citep{Zhu2019,hau20,bianco2021,bekki2021}. Most previous methods are based on classical machine-learning methods, such as Support-Vector Machine (SVM) or Random Forest~\citep{hastie2001elements}. These methods must extract an adapted set of features in order to be sufficiently efficient: we have to be sure that the extracted features represent the subject we wish to study well. These features are usually given by expert knowledge of the problem.  

Most successful machine-learning methods for image processing tasks today are based on deep neural network methods~\citep{goodfellow2016deep}. These methods have the particularity of learning both the task and a representation of the data dedicated to the task itself. Thus, they are more powerful than methods based on hand-tuned features. While the first methods were dedicated to classification (e.g., AlexNet \citep{Krizhevsky2012ImageNetCW}, LeNet \citep{LeCun1989BackpropagationAT}, or ResNets \citep{ResNetHe2016}), there are now many different architectures depending on the targeted task. For segmentation, one of the most promising neural networks is the UNet, which was introduced for medical segmentation~\citep{ronneberger2015u}. Many extensions exist, for instance, UNet++~\citep{zhou2019unet++} with layers to encode the concatenations, VNet~\citep{milletari2016v}, which is dedicated to 3D data, WNet~\citep{xia2017w}, which has a double UNet architecture, and Attention-UNet~\citep{oktay2018attention}, which combines UNet with attention layers~\citep{goodfellow2016deep}.
Still, the UNet based architecture remains one of the most  effective methods for automatic segmentation.

In the context of astrophysical study, these neural networks have been successfully used in different contexts. For example, \citet{bekki2021} used the UNet to segment the spiral arms of galaxies.~\citet{bianco2021} used a UNet based neural network called SegUNet to identify H\,{\sc{ii}} regions during reionization in 21 cm.  Another variant based on UNet and inception neural networks was used to predict localized primordial star formation~\citep{wells2021predicting}. UNet was also used to segment cosmological filaments~\citep{aragon2019}. We recommend~\citet{hau20} for a good introduction to deep learning applied to astrophysical data.
\begin{figure}[htb]
   %\centering
\includegraphics[scale=0.35]{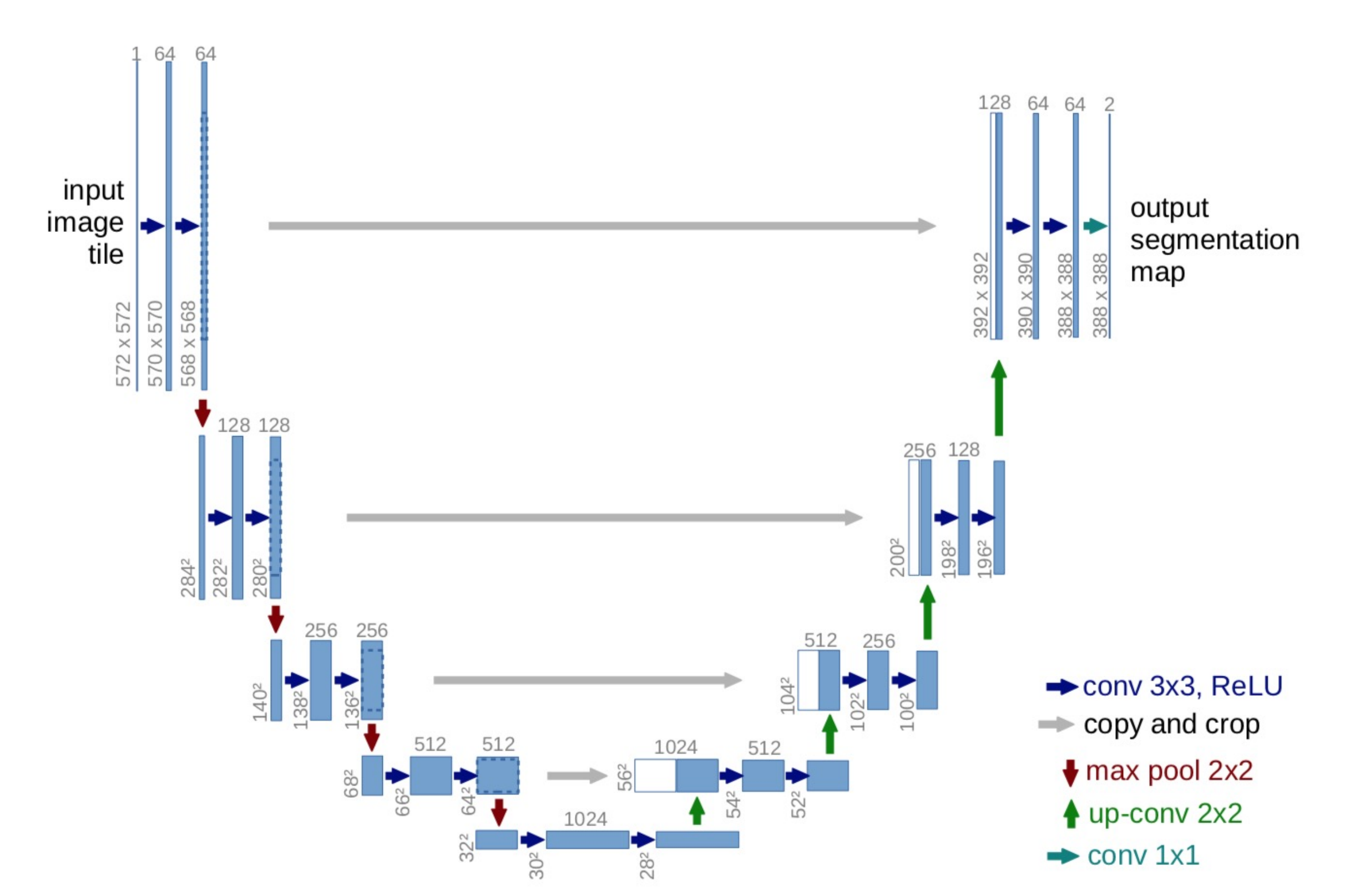}
   \caption{Illustration of the UNet5 from~\citet{ronneberger2015u}.}
        \label{UNet}%
\end{figure}

UNet is a multiscale neural network based on convolutional and pooling layers, as presented in Figure~\ref{UNet}. In addition to its simple structure, the strength of this network is an encoder-decoder-based architecture with skip connections. First, the encoder extracts features from the input image down to a coarse scale by using filters and max-pooling. Then, the decoder takes the coefficients at the coarse level and combines them with those from each layer of the encoder via the skip connections, in order to re-inject the details that were lost in the down-sampling (max-pool) step and thereby build a better semantic segmentation map. The final activation function of the network is done by the sigmoid function~\citep{goodfellow2016deep}, as we wish to have values in order to resolve a segmentation issue,
\begin{equation}
    s(x) = \frac{\exp(x)}{\exp(x) + 1}~.
\end{equation}
This function guarantees an output between 0 and 1. Thus, the output of the network can be read as a probability map for the class 1 filament~\citep[see][Sect.6.2.2.2]{goodfellow2016deep}. However, in our case, both the nonequilibrium between the two classes (filament and background) and the incomplete ground-truth prevent the direct interpretation of the segmented map values as probabilities~\citep[see][about sigmoid output and probabilities]{kull2017beyond}. In the following, we name the intensity value of the segmented maps "classification value". In this study, the quality of the results is assessed by comparing these classification thresholds with a given threshold (see Sect.~\ref{metrics}).
This multiscale mirror-like structure makes the UNet very suitable for image processing such as denoising~\citep{batson2019noise2self} or segmentation~\citep{ronneberger2015u}. Moreover, UNet belongs to the family of fully convolutional networks. These networks are almost independent of the size of the input images~\citep{long2015fully}. In UNet, the size of the output image will be the same as that of the input if the input is large enough (the minimum size is $32\times 32$ pixels).

\begin{figure}[htb]
   \centering
\includegraphics[width=0.9\linewidth]{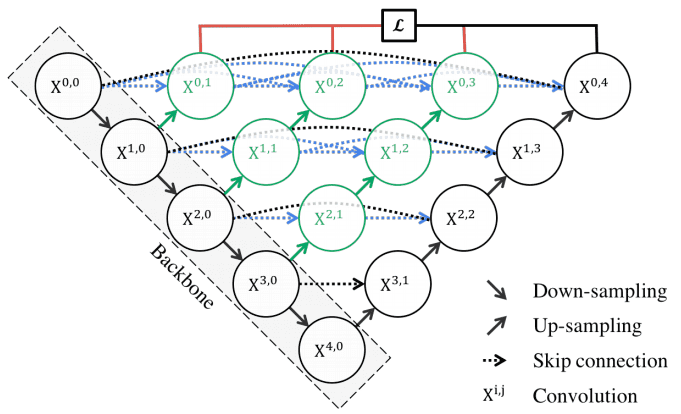}
   \caption{Illustration of UNet++ from~\citet{zhou2019unet++}. The $X^{i,j}$ are the same convolutional layers as for UNet. The difference between UNet and UNet++ can be depicted in three main points: 1) convolution layers on skip pathways (in green), which reduces the semantic gap between encoder and decoder feature maps; 2) dense skip connections on skip pathways (in blue), which improves the gradient flow; and 3) deep supervision (in red), which enables model pruning ~\citep{pmlr-v38-lee15a}.}
        \label{fig:UNetpp}%
\end{figure}

A recent and more powerful extension of the UNet model, UNet++, was proposed in ~\citep{zhou2019unet++}. This network belongs also to the fully convolutional networks. As illustrated in Fig.~\ref{fig:UNetpp}, the plain skip connections of UNet are replaced with a series of nested dense skip pathways in the UNet++ neural network. The new design aims at reducing the semantic gap between the feature maps of the encoder and decoder sub-networks that makes the learning task easier to solve for the optimizer. In fact, the model captures more efficiently fine-grained details when high-resolution feature maps from the encoder are gradually enriched before fusion with the corresponding semantically rich feature maps from the decoder. Note that these "inner" layers have also a mirror-like structure allowing a larger multi-scale representation. However, it is worthy to note that UNet++ requires more data than UNet as the latter has less parameters to tune \citep[see Table 3 in][]{zhou2019unet++}.
\subsubsection{Local normalization}
\begin{figure} \label{norm}
    \centering
    \includegraphics[width=0.8\linewidth]{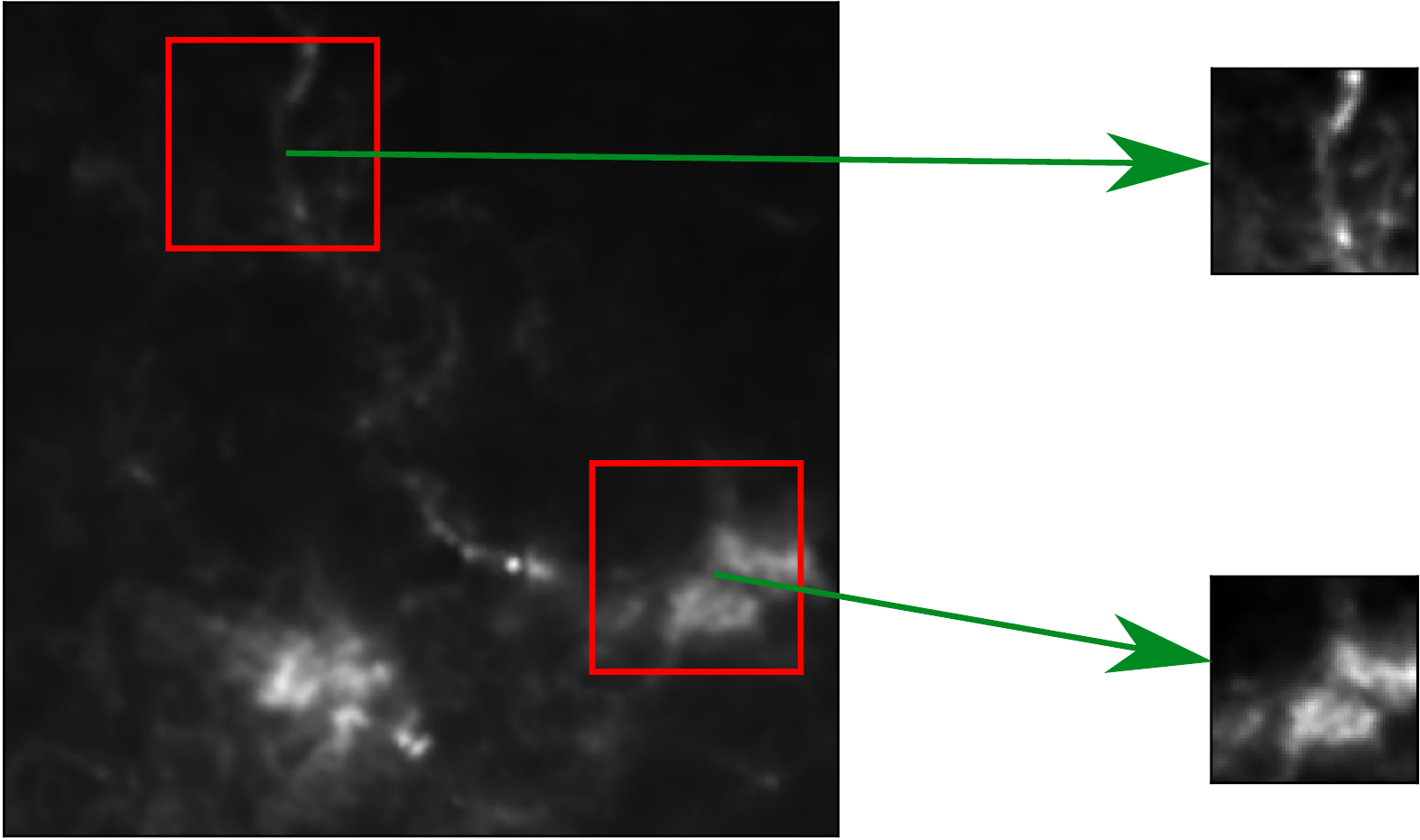}
    \caption{The \textit{local} min-max normalization of the patches helps to avoid contrast issue allowing a better definition of the filaments. }
    \label{fig:normalization}
\end{figure}

Neural networks such as UNet are highly sensitive to the contrast inside the input images (or patches). This sensitivity comes from the filters that belong to the different convolution layers. In order to avoid this issue, input data are usually normalized, generally by performing a global min-max normalization~\citep{goodfellow2016deep}. This normalization allows us to temper the dynamic of the contrast while keeping useful physical information about the structures (morphology and gradient).
However, in our case, the intensity of the N$_{\rm{H_2}}$ map presents a very high dynamical range, and a global normalization would artificially weaken many filamentary structures. To avoid this issue, we performed a \textup{\textit{local}} min-max normalization on each patch. As shown in Figure~\ref{fig:normalization}, this normalization helps to deal with high-contrast variation in nearby regions of the image. However, while this approach solves this issue, the contrast still has high local variations in some cases, so that two nearby patches may show different normalization. 

\subsubsection{Training with UNet and UNet++}
Training a neural network requires a loss function that computes the errors between the model and the ground truth. For segmentation, a recommended function is the binary cross-entropy (BCE), which casts the problem as a classification problem \citep{Jadon2020ASO}. 
For the sake of clarity, we introduce some notations before we give the expression of the loss function. Let $\{x_i\}_i$ be the set of normalized N$_{\rm{H_2}}$ patches. Let $\{y_i\}_i$ be the set of segmentation target, that is, the set of binary patches with 1 for filaments pixels and 0 for background pixels. Let $\{m_i\}_i$ be the set of missing data patches, that is, the set of binary patches with 0 for missing pixels and 1 elsewhere. For a given value of $i$, $x_i, y_i$, and $m_i$ share the same Galactic coordinates.
The cross-entropy loss for a set of $n$ patches is given by
\begin{equation}
    \begin{split}
        \mathcal{L}(\{x_i, y_i, m_i\}_i ; \theta) = \frac{1}{np^2} &\sum_{i=1}^n \sum_{k,l=1}^p m_i[k, l] \Big(y_i[k,l] \log(f_\theta(x_i)[k,l])+\\ &(1 - y_i[k,l])\log(1-f_\theta(x_i)[k,l]) \Big)
    \end{split}\label{eq:loss}
,\end{equation}
where $f_\theta$ is the function that applies the forward propagation, and $\theta$ are the weights of the neural network. By using $\{m_i\}_{i\in[i\ldots n]}$, we ensure that only labeled data are used.

\begin{figure*}[ht]
    \includegraphics[width=0.95\linewidth]{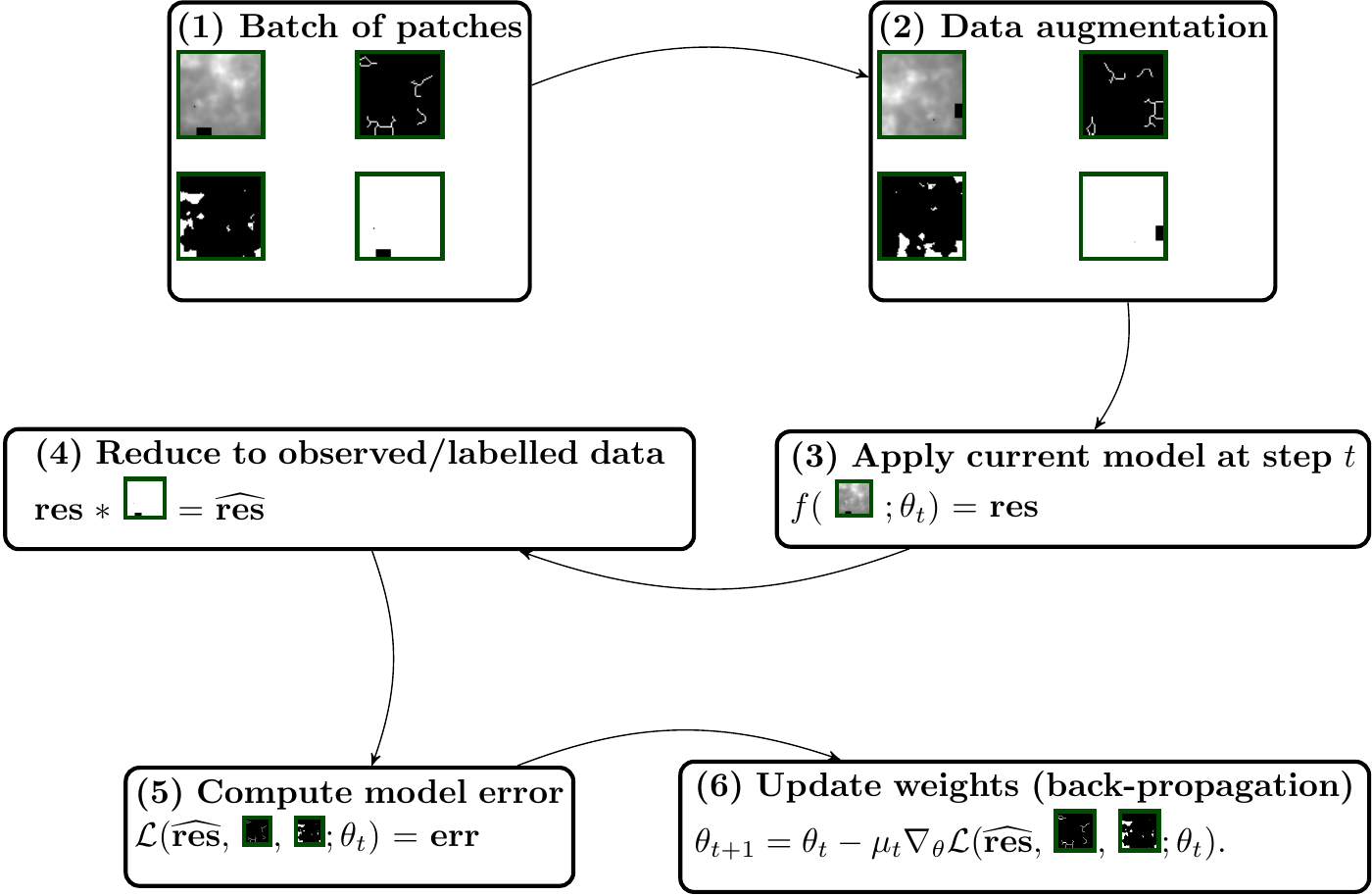}
    \caption{Five steps of an epoch during the training. For illustration purposes, we reduced the batch to  a set of one patch. $\theta_t$ represents the weights of the neural network at epoch $t,$ and $\mu_t$ is the learning rate at epoch $t$.}
    \label{fig:learning_proc}
\end{figure*}

As we have many unlabeled pixels in the patches (see section~\ref{datasetbuilding}), we have to adapt the training step to avoid inconsistencies. We summarize the different steps in Figure~\ref{fig:learning_proc}. First, in step (1), we take a set of patches (a batch) and then apply the augmentation process (step (2)) on these patches. During this step (2), we ensure that for a given $i$, the same transformation is applied on $x_i, y_i$, and $m_i$. The following steps are about computing the prediction errors of the model on the patches and making the back-propagation of the gradient of these errors in order to update the weights of the neural networks ~\citep{goodfellow2016deep}. Therefore, in step (3), we begin to apply the network $f_{\theta_t}$ on the patches (forward propagation). Since this step implies using the convolution layers in the network, we use both unlabeled and labeled pixels. This is important as the neural network needs the neighboring pixels to compute the value for one pixel. After we restrict the result (step (4)) to labeled and nonmissing pixels using the mask $m_i$, we can compute (step (5)) the errors on the restricted results compared to the ground truth. Finally, in step (6), we update the weight of the network using the back-propagation of the gradients of the errors. This is done by using a stochastic gradient descent scheme~\citep{goodfellow2016deep} with a learning rate $\mu_t$ that changes during the training step following the epochs. 
\subsubsection{Building the segmentation map} \label{seg}
\begin{figure}[ht]
    \includegraphics[width=0.8\linewidth]{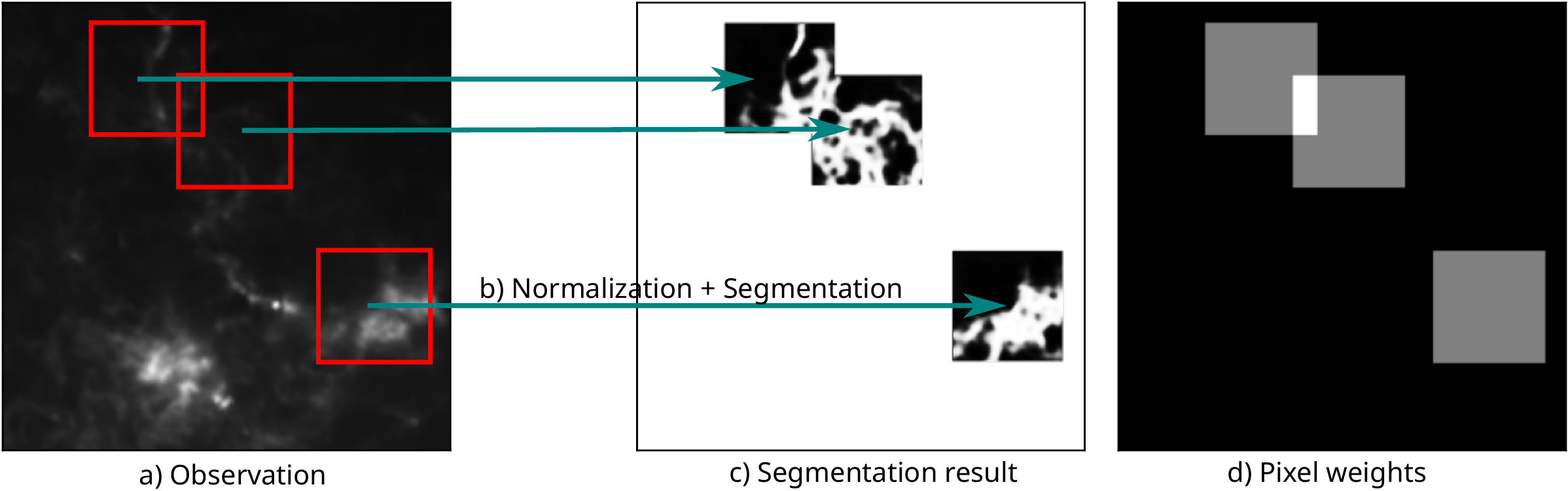}
    \caption{Segmentation process. It takes patches from an observation (a), then normalizes the patch and applies the segmentation model (b), the segmented patch is positioned at the same coordinates (c), and is finally weighted by coefficients (d) representing the number of patches
    in which each pixel appears. Because a sliding window with overlap is used, a given pixel is segmented several times (as long as it falls in the sliding window). Then, we obtain several segmentation values for the same pixel. The final segmentation value assigned to the pixel corresponds to the average of all the segmentation values computed from the contributing sliding windows.}
    \label{fig:segmentation_process}
\end{figure}

When the neural network has been trained, we can apply the model to segment an image. As we described before, during the creation of the data set (Section~\ref{datasetbuilding}), we must deal with the high dynamic contrast in images. Again we propose to solve the issue by taking small patches and apply a local min-max normalization. Moreover, since two closed patches may have a different contrast, the normalization can lead to variation in the results when applying the neural network. Therefore, in order to resolve this issue, we propose to use an overlapping sliding window to obtain the patches: the segmentation result is then the average between the output of the neural network applied on the patches. These patches are distinct from those used for learning the network (see Section \ref{datasetbuilding}). As the variance of contrast between patches introduced variance inside the output results of the neural network, the overlap and averaging operation (see Fig.\ref{fig:segmentation_process}) allows us to decrease the artifact that may appear~\citep{pielawski2020introducing}.

Thus, we apply the following segmentation procedure (illustrated in Figure~\ref{fig:segmentation_process}). We browse the image using an overlapping sliding window that gives patches (step (a)). Each patch is then normalized using a min-max normalization (step (b)); here we avoid the missing data (around borders and saturated areas). Then, we apply the trained neural network on the patch to obtain the density map output and add on the output image at the coordinate of the patches (step (c)). Since we use an overlapping sliding window, the results are added to the output, and then we divide each pixel by a weight representing the number of patches in which the pixel appears (this is done using the weight map built in step (d)).

\subsubsection{Metrics}\label{metrics}
In supervised classification problems, the confusion matrix, also called error matrix, is computed in order to assess the performance of the algorithm. We refer to the filament and background classes as the positive (P) and the negative (N) classes, respectively. We also refer to the correctly and misclassified pixels as true (T) and false (F), respectively. In a binary classification problem, the confusion matrix is thus expressed as in Table~\ref{tab:ConfusionMatrix}.

\begin{table}
\caption{Confusion matrix} \label{tab:ConfusionMatrix}
\begin{tabular}{cc|cc}
\multicolumn{2}{c}{}
            &   \multicolumn{2}{c}{Predicted} \\
    &       &   filament &   background              \\ 
    \cline{2-4}
\multirow{2}{*}{\rotatebox[origin=c]{90}{Actual}}
    & filament   & \rule{0em}{1.5em} TP   & FN                 \\
    & background    & \rule{0em}{1.5em} FP    & TN                \\ 
\end{tabular}
\end{table}
The confusion matrix is evaluated on the estimated filament masks that are deduced from the segmented map at a classification threshold. It is important to recall, however, that the evaluation set is restricted to labeled data. The classification scores are thereafter derived from the confusion matrix. In this work, the recall, precision, and dice index defined in \ref{eq:scores1}, \ref{eq:scores2} and \ref{eq:scores3}, respectively, are used to evaluate the classifier performance. Maximizing recall and precision amounts to minimizing false-negative and false-positive errors, respectively, whereas maximizing the dice index amounts to finding the optimal tradeoff between the two errors. Therefore, the closer to 1 these scores, the better.
\begin{align} \label{eq:scores1}
 \mathrm{Recall} &= \frac{TP}{TP + FN} \\ \label{eq:scores2}
 \mathrm{Precision} &= \frac{TP}{TP + FP} \\ 
 \mathrm{Dice} &= \frac{2\ TP}{2\ TP + FP + FN} \label{eq:scores3}
\end{align}
In addition to the recovery scores, we also calculate the rate of missed structures (MS) in segmentation for the same set of thresholds. The MS score can be defined as the ratio of the missed filament structures over all the input filament structures,

\begin{equation}\label{eq:MS}
\mathrm{MS} =\frac{\text{number of missed structures}}{\text{total number of filament structures}}
\end{equation}

\begin{figure*}[ht]
    \includegraphics[width=0.7\linewidth]{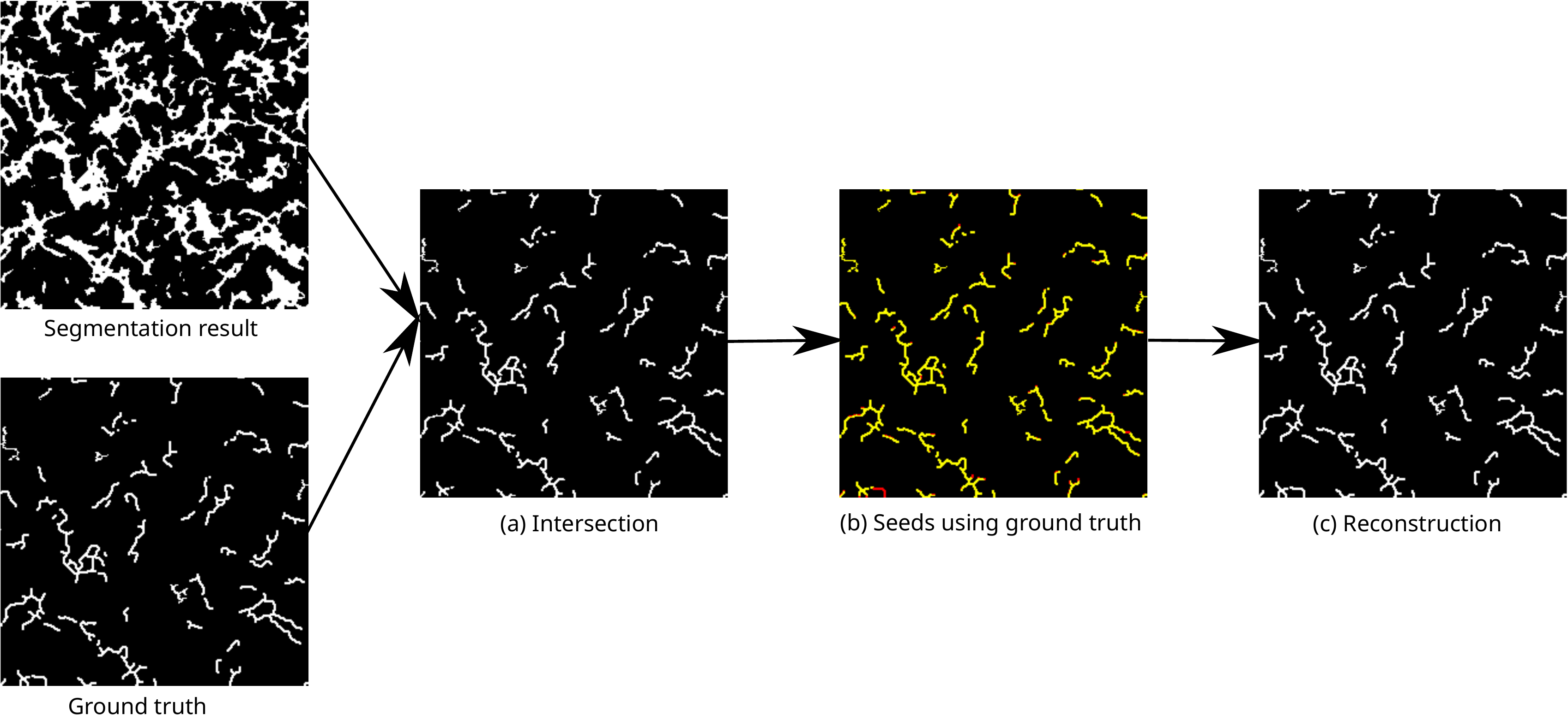}
    \caption{Morphological reconstruction is a method that computes shapes from marked pixels called seeds. (a) We first compute the seeds using the intersection between the segmentation results and the ground truth. (b) We use the intersection pixels as seeds (see the red seeds in the bottom left corner). (c) We apply the reconstruction to obtain the filaments with at least one seed.}
    \label{fig:reconstruction}
\end{figure*}
The MS metric is based on morphological reconstruction~\citep{Vincent93,Soille2003,Robinson2004}. Figure~\ref{fig:reconstruction} shows how we apply this method to assess which known filaments are recovered. First, we compute the intersection between the known filaments (ground truth) and the segmentation results (Figure~\ref{fig:reconstruction}(a)). Then we use this intersection as the seed for the reconstruction method: in Figure~\ref{fig:reconstruction}(b), the yellow elements are the seeds, and the red elements are the part of filaments that is missed. The morphological reconstruction takes the seeds to recover the known filaments by using a shape-constrained growing process (Figure~\ref{fig:reconstruction}(c)). Only the filaments for which at least one pixel is used as seed will be recovered~\citep{Robinson2004}.  By subtracting the morphological reconstruction result from the ground truth, we can identify the missed structures (see also Figure~\ref{missed}). Then we count the number of missed structures by using a direct labeling of the pixels where two neighboring pixels share the same label~\citep{fiorio1996two}. Then, we can compute the MS metric and qualitatively assess the recovery of filaments in terms of structures, rather than individual pixels. \textbf{}  

\subsection{Experimental setup}\label{setup}
Taking the analysis of the normalization in Section~\ref{datasetbuilding} into consideration, the patch size was fixed to the lowest value accepted by the UNet family, $p = 32$. In order to have proper training, validation, and test steps, we randomly split our initial set of patches (Section~\ref{datasetbuilding}) into three sets, namely, the training, validation, and test sets, with proportions of  $80\%$, $10\%,$ and $10\%$, respectively. The random split ensures the presence of the two classes (filament and background) in the three sets. The patches in training and validation sets were then shuffled after each epoch to help avoiding unwanted bias \citep[see]{goodfellow2016deep}. The total number of epochs was set to $100$. 
 UNet and UNet++ were trained using the Adam optimization scheme with a multistep learning rate~\citep{kingma2015adam, ronneberger2015u}. During the first 30 epochs, the initial learning rate value was divided by $10$ every $\text{five}$ epochs. Four initial values of the learning rate, $10^{-5}, 10^{-4}, 10^{-3}, \text{and }10^{-2}$, were tested for both networks. We  denote by UNet[$\mathrm{lr}$] (UNet++[$\mathrm{lr}$]) the UNet (UNet++) model learned with $\mathrm{lr}$ as the initial value of the learning rate, where $\mathrm{lr} \in \{10^{-5}, 10^{-4}, 10^{-3}, 10^{-2}\}$. A summary of the parameter values used in the training step is given in Table \ref{tbl:ExpSetup}. 
In the segmentation step, an overlapping sliding window of size $32 \times 32$ was applied, where an overlap of $30$ pixels was used in order to limit edge artifacts and to generate highly smooth segmentation.

In order to compare the performance of UNet with UNet++, two zones of the global N$_{\rm{H_2}}$ mosaic were excluded from the initial patch data set. Constrained by the limited number of patches in the data set, small zones were removed, namely, the zones 166.1-168.3\textdegree{} and 350.3-353.5\textdegree. The choice of the removed regions was motivated by assessing the network performance in regions with a highly diverse column density and filaments content. While 350.3-353.5\textdegree, removed from mosaic 349-356\textdegree, is dense and rich in filaments ($38541$ filament pixels), the region 166.1-168.3\textdegree, removed from the mosaic 160-171\textdegree, is sparser and contains fewer filaments ($1459$ filament pixels). The filament density is always inferred based only on the incomplete ground truth (labeled part of the data set). The two removed zones were then segmented by the learned models, and the segmentation quality was assessed using the evaluation scores described in Section~\ref{metrics}. The evaluation scores were also computed on the fully segmented Galactic plane in order to have a global performance evaluation of models. 

\begin{table}[ht]
      \caption{Experimental setup}
    \label{tbl:ExpSetup}
    \begin{tabular}{l@{\rule{3em}{0em}}l}
         \hline \textbf{Parameter} & \textbf{Value} \\
         \hline patch size ($p$) & 32 pixels \\
         dataset split & \{80\%, 10\%, 10\%\} \\
         batch size & 64 patches \\
         epochs & 100 \\
         initial learning rate & \{$10^{-5}, 10^{-4}, 10^{-3}, 10^{-2}$\} \\
         \hline     
    \end{tabular}
    \tablefoot{Parameters used in the experimental setup for UNet and UNet++ training}   
\end{table}

\section{Results}\label{res}
\subsection{Scores and segmented mosaic analysis}\label{sec:ScoresAnalysis}
In order to evaluate the training performances, we discuss below the different scores we obtained for the different tested scenarios. Two neural networks were tested (UNet and UNet++) with four different initial learning rates for each.  For an input column density image, the segmentation process (see Section~\ref{seg}) returns a classification value mask from which it is possible to identify pixels that likely belong to a filamentary structure. Nevertheless, we did not attempt to extract the filaments like \citet{schisano2020} did. We postpone this physical analysis on the newly identified filaments to a follow-up work. Here we present the method as a proof-of-concept and analyze its performances and returned results (segmented map of the whole Galactic plane). We illustrate these results on two portions of the Galactic plane that were selected for their characteristics in terms of column density and filament content (as known from the input data set of filament mask based on the spine+branches).   \\
To evaluate the performances of the different scenarios, Figure~\ref{fig:BCETrainVal} presents the BCE curves for the  training set and the validation set during the first $30$ epochs. For all models, these loss curves reach a plateau around the $\text{tenth}^{}$ epoch, confirming the rapid convergence of UNet-based networks. The performance of UNet++[$10^{-2}$] was removed from the displayed results due to convergence issues. As shown in Figure~\ref{fig:BCETrainVal} and confirmed by this analysis, the models have similar performances and converge to a training error around $0.01$, except for schemes with a starting learning-rate value of $10^{-5}$ , where higher errors are reported (more than $0.016$ and $0.012$ for UNet[$10^{-5}$] and UNet++[$10^{-5}$], respectively). The similar performances obtained for the two architectures and the different learning rates indicate that the method is robust. Validation errors are slightly higher than training errors, with a difference up to $0.0011$, except for UNet[$10^{-5}$] (around $0.002$) and UNet++[$10^{-5}$] (around $0.0018$). The reported values indicate that the learned models show low bias and low variance. \\
For both UNet and UNet++, the best performance is with an initial learning-rate value of $10^{-3}$. Overall, the best model in terms of loss function is the UNet++[$10^{-3}$]. For each scenario, the best-performing model in validation was used to segment the test set and compute the underlying BCE (see Table~\ref{tbl:BCE-test}). BCEs calculated on the test set corroborate the previous analysis: scenarios with initial learning rates in [$10^{-2}$, $10^{-3}$, $10^{-4}$] result in similar test errors that do not exceed $0.009$, whereas scenarios with initial learning rates of $10^{-5}$ show a higher error. The lowest test error corresponds to the scenario UNet++[$10^{-3}$]. \\
\begin{figure*}[h]
\includegraphics[width=\linewidth]{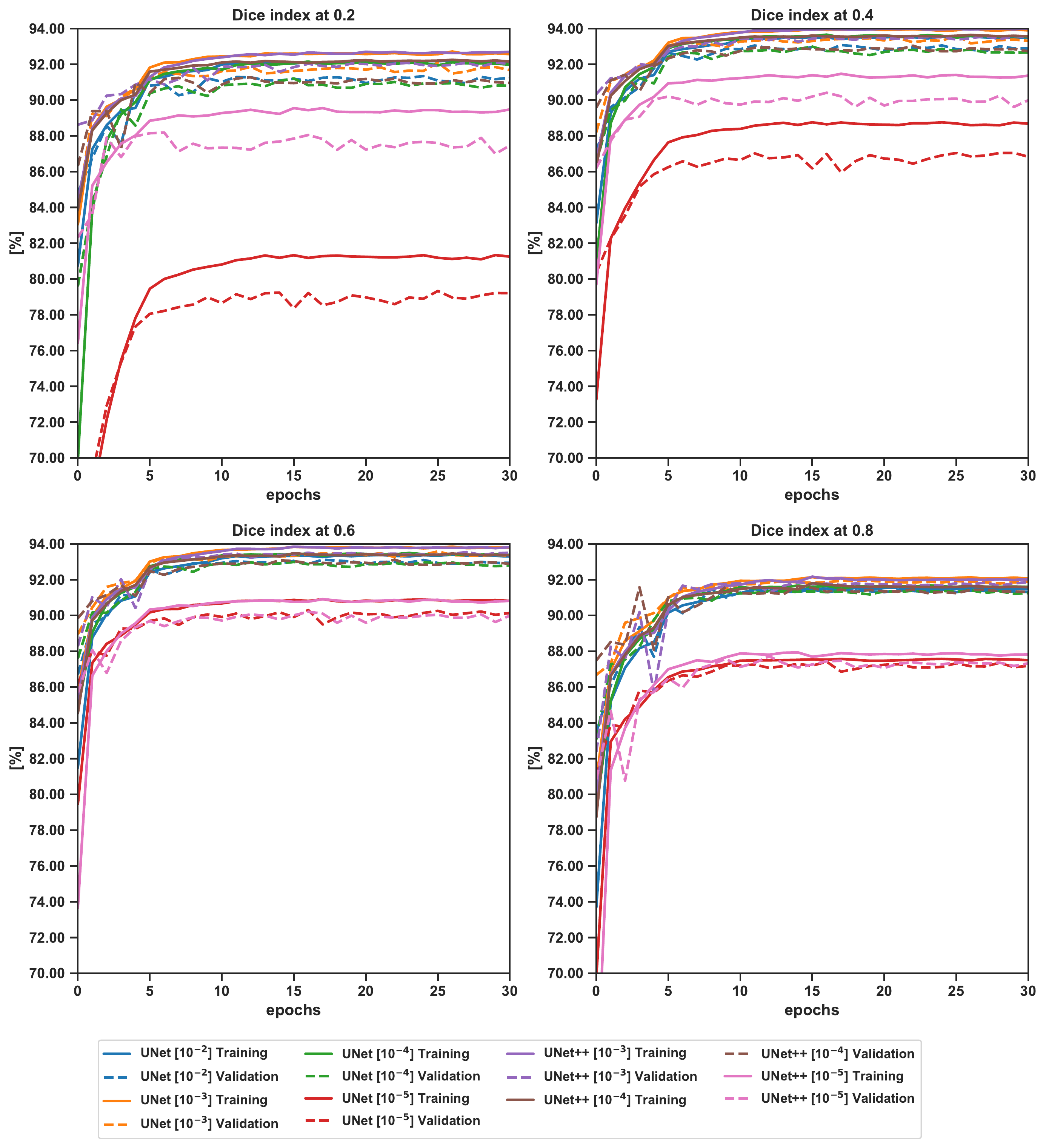}
   \caption{Dice curve evolution of schemes reported in Figure \ref{fig:BCETrainVal} are displayed over the first $30$ epochs in training (continued lines) and validation (dashed lines) steps, at classification threshold values $0.8$, $0.6$, $0.4,$ and $0.2$. The displayed curves are aligned with the results deduced from BCE curves, where close performances were obtained with initial learning-rate values of $10^{-4},10^{-3},\text{and }10^{-2},$ and a poorer performance is obtained for schemes with a learning-rate value of $10^{-5}$. The highest dice score is reported for UNet++[$10^{-3}$] (purple), and the lowest performance corresponds to the UNet[$10^{-5}$] scheme, (red) especially at thresholds 0.2 and 0.4. Similarly to the BCE curves and for all displayed schemes, a plateau regime is reached within the first $\text{ten}$ epochs.}
              \label{fig:DiceTrainVal}
    \end{figure*}
In Figure~\ref{fig:DiceTrainVal} the dice index curves for the  training and validation sets is plotted for different classification thresholds. The dice index was also computed on a test set using the best-performing models in validation (the models used in Table \ref{tbl:BCE-test}), and the corresponding results are given in Figure~\ref{tbl:Dice-test}. In the same way, the dice index results are in line with the BCE. These models were used afterwards to segment the removed zones 166.1-168.3\textdegree{} and 350.3-353.5\textdegree.
\begin{figure}[t]
   \centering
\includegraphics[width=\linewidth]{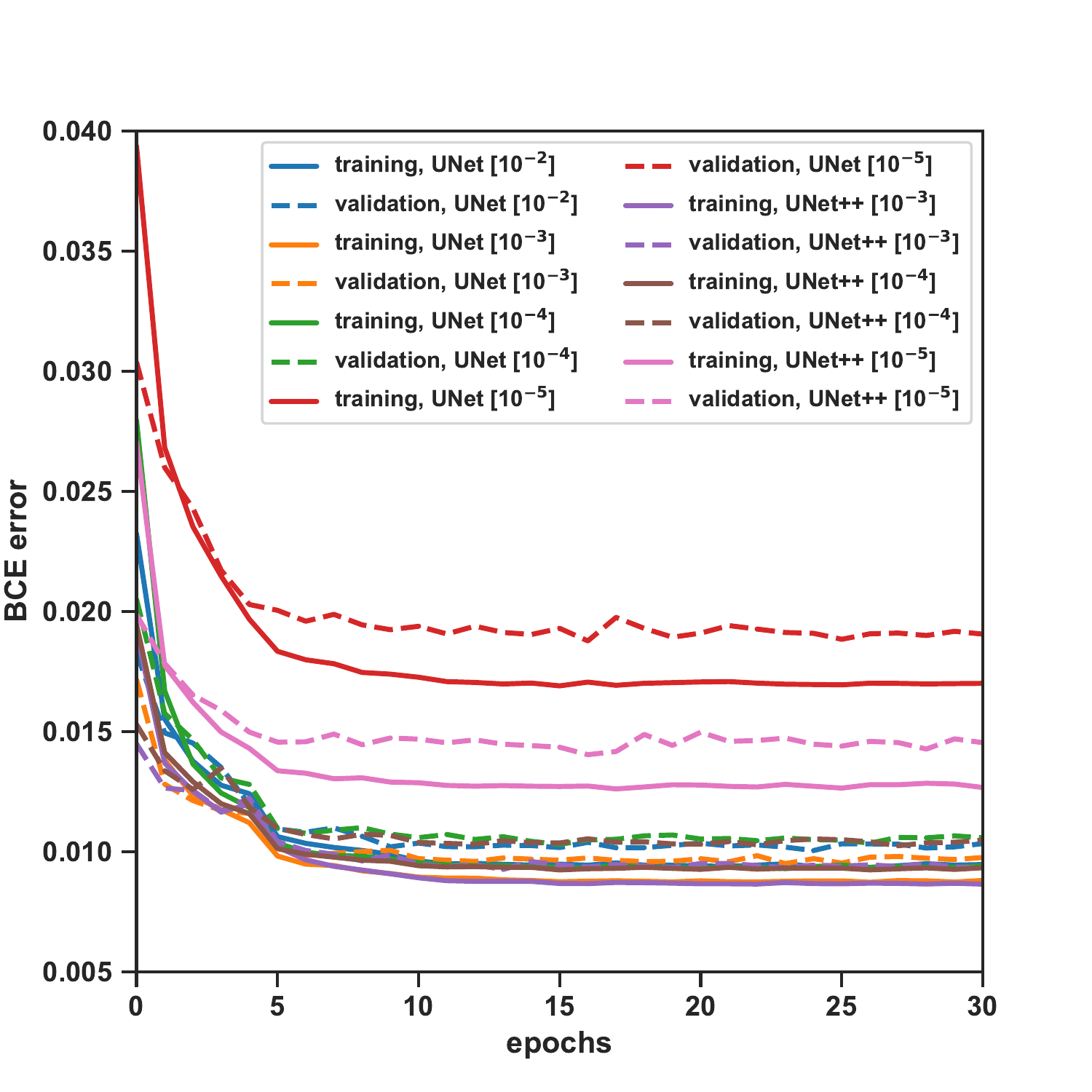}
   \caption{BCE evolution over the first $30$ epochs in training (continued lines) and validation (dashed lines) steps. UNet++[$10^{-2}$] schemes is removed as its corresponding BCE diverged. The displayed schemes show similar performances. Models with a learning-rate value of $10^{-5}$ resulted in higher BCEs. The lowest error is reported for UNet++[$10^{-3}$] (purple), and the highest error corresponds to the UNet[$10^{-5}$] scheme (red). A  plateau regime is reached within the first $\text{ten}$ epochs for all models, which confirms the rapid convergence of the UNet-based networks.}
    \label{fig:BCETrainVal}
\end{figure} 
 Figure~\ref{res-mods} presents the results of the segmentation process for the different scenarios tested and presented in Table~\ref{tbl:ExpSetup}. All segmented maps are presented within the range $[0,1]$. Overall, the obtained images are in line with the analyzed BCEs. The scenarios with initial learning rates of $10^{-2}$, $10^{-3}$, $\text{and }10^{-4}$ return very similar segmented maps. A noticeable difference is seen in maps that were segmented by both UNet and UNet++ with an initial learning rate of 10$^{-5}$. In these cases, the filamentary structures are broader than those obtained for the other scenarios, especially for UNet. Moreover, the intensity of the segmentation maps also varies for these last two scenarios, where the low-value structures are better revealed (intensity variation up to a factor of 10 in those zones). The ability of these two last scenarios (and of the UNet in particular) to better reveal structures with a lower classification threshold might be used to detect structures that are not well seen on the original map, either due to their low contrast and/or their low column density. A close visual inspection of the column density images confirms that features revealed by UNet[$10^{-5}$] and UNet++[$10^{-5}$] are low-contrast filaments that were present in the original images, but absent from our input catalog of filaments that was used as ground truth. \\
\begin{figure*}[t]
\begin{subfigure}{.5\textwidth}
  \centering
  \includegraphics[width=\linewidth]{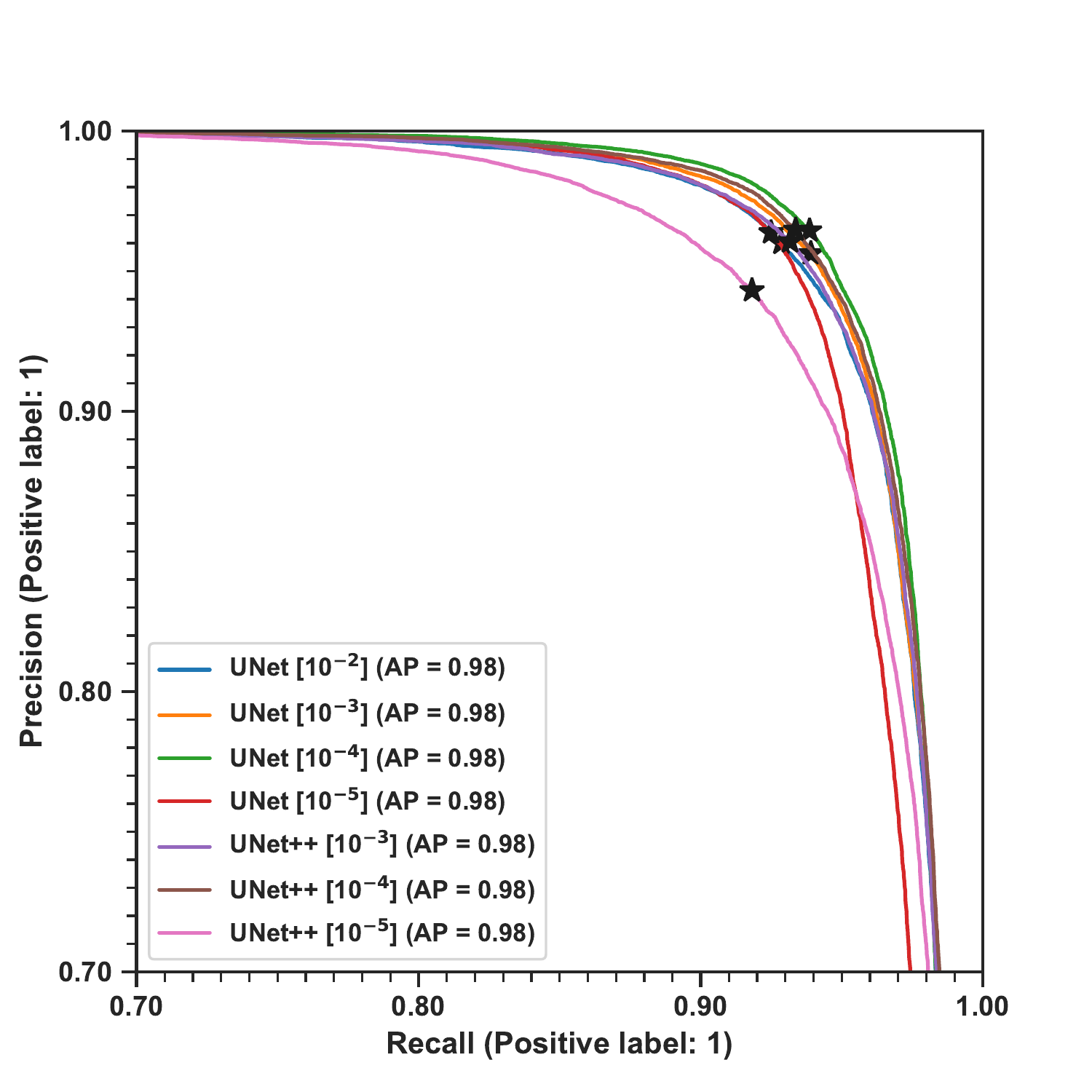}  
  \caption{PR curves for 350.3-353.5\textdegree}
  \label{fig:PR-a}
\end{subfigure}
\begin{subfigure}{.5\textwidth}
  \centering
  \includegraphics[width=\linewidth]{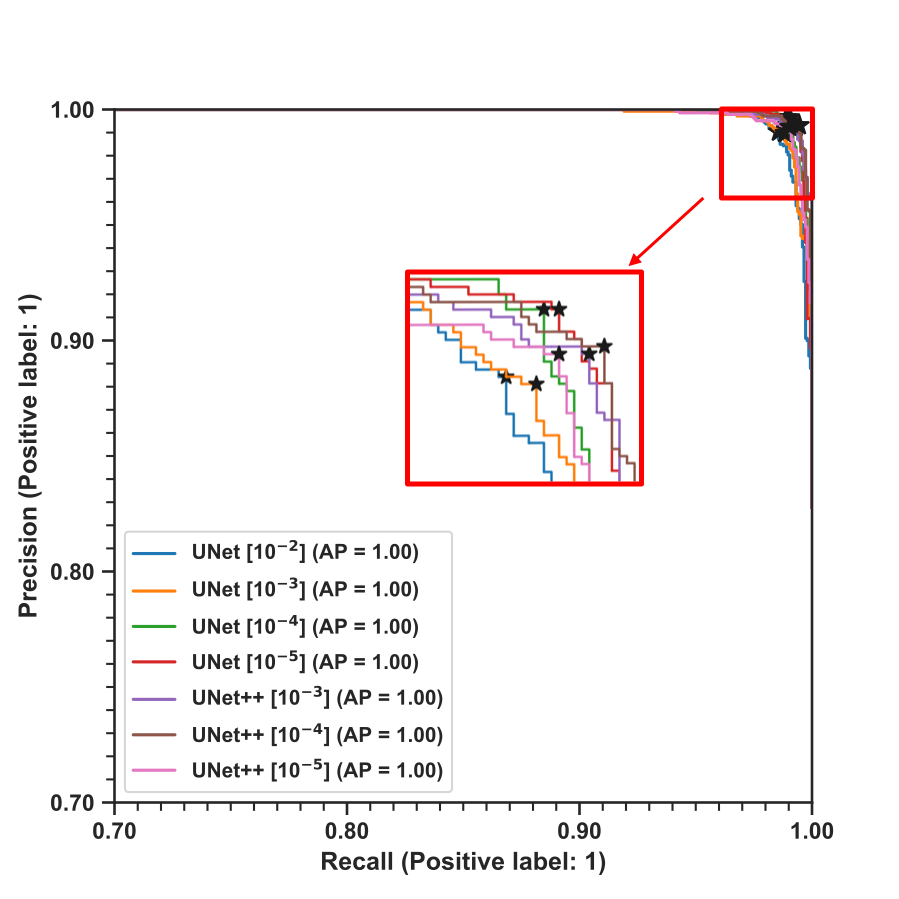}
  \caption{PR curves for 166.1-168.3\textdegree}
  \label{fig:PR-b}
\end{subfigure}
\newline
\begin{subfigure}{.5\textwidth}
  \centering
  % include third image
  \includegraphics[width=\linewidth]{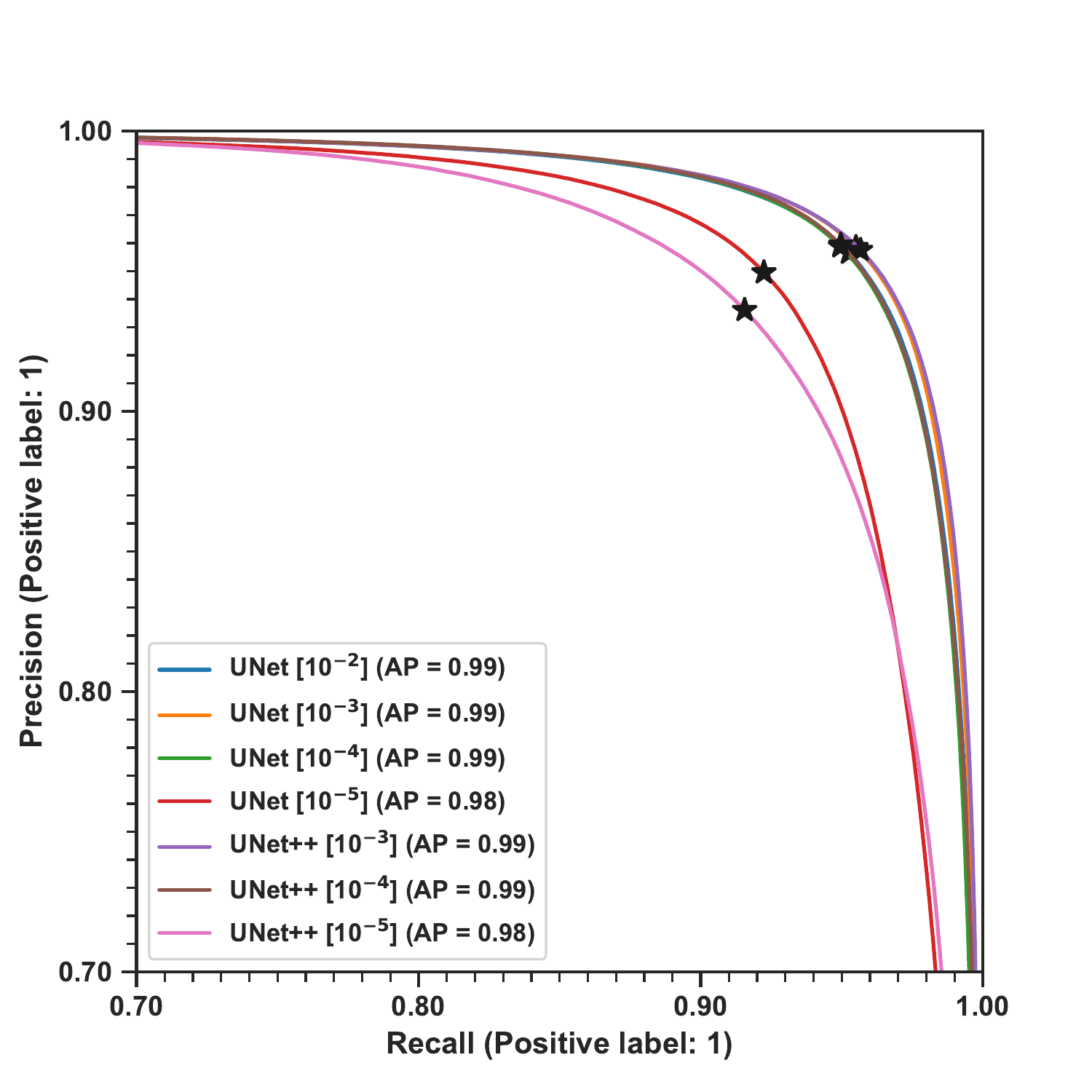}  
  \caption{PR curves for the fully segmented Galactic plane}
  \label{fig:PR-c}
\end{subfigure}
\caption{P-R curves of the schemes reported in Figure \ref{fig:BCETrainVal}, computed on the segmented removed zones (top) and the full Galactic plane (bottom). (a) P-R curves computed on the segmented 350.3-353.5\textdegree{} , which corresponds to the dense region that was removed from the patches data set. (b) P-R curves computed on the segmented 166.1-168.3\textdegree{} , which corresponds to the sparse region that we removed from the patches data set. (c) P-R curves computed on the full segmented Galactic plane. Unlike in Fig. \ref{fig:PR-b}, P-R curves obtained on the latter are close to those obtained in Figure \ref{fig:PR-a}.}
\label{fig:PR-curves}
\end{figure*}
\begin{figure*}[htb]
   %\centering
\includegraphics[width=\linewidth]{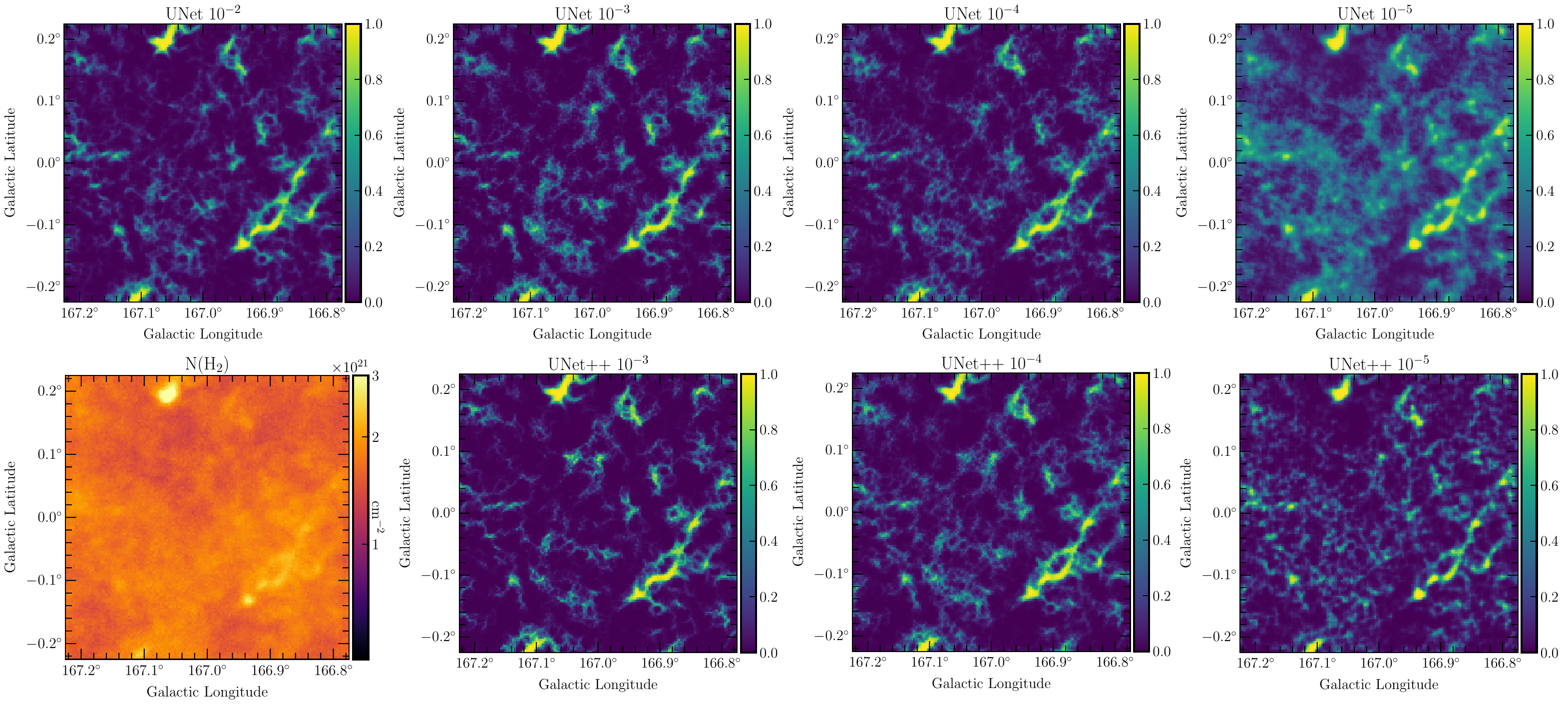}
\includegraphics[width=\linewidth]{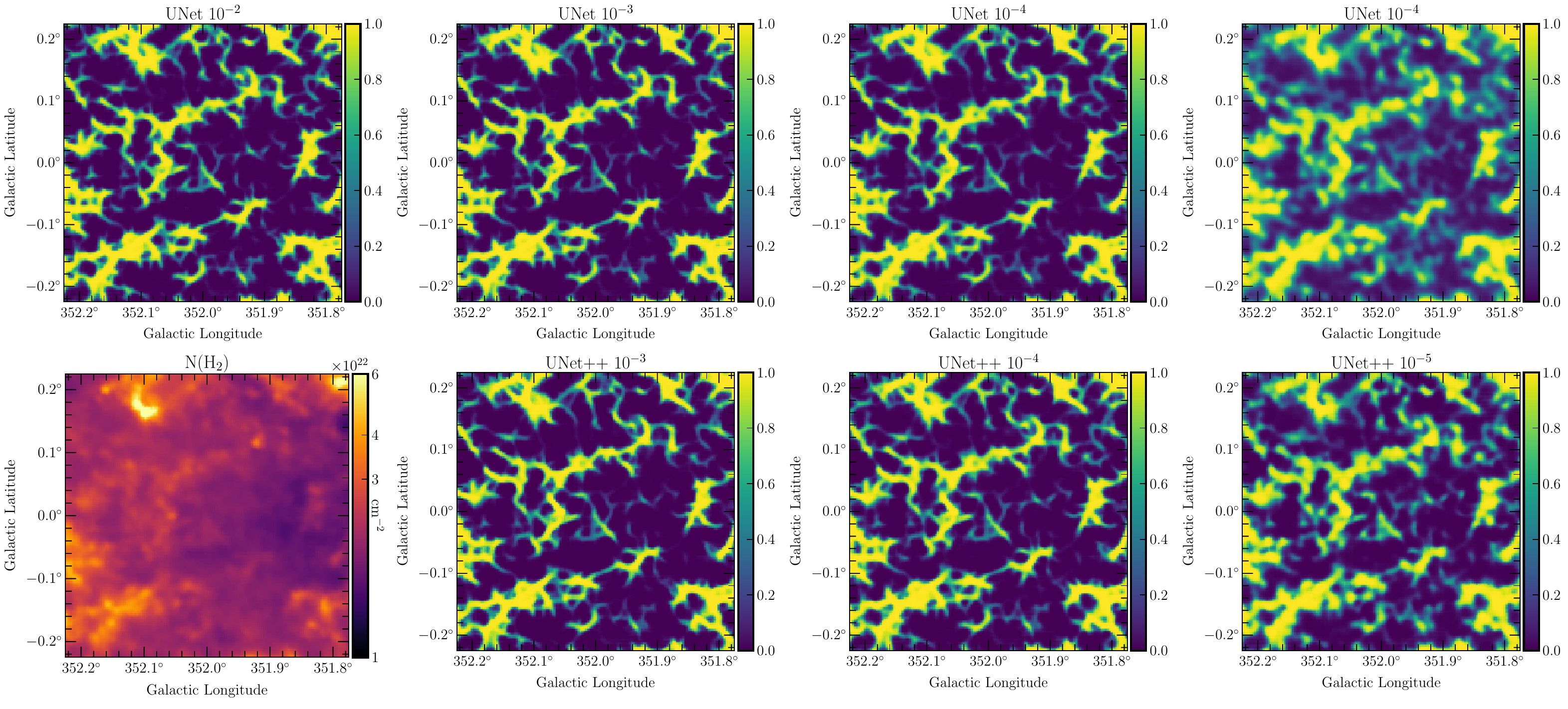}
   \caption{Segmented maps obtained for the models analyzed in Figure~\ref{fig:BCETrainVal}, zooming in on a part of the two regions that was removed from the training namely, the $l$=166.1-168.3\textdegree{} (top) and $l$=350.3-353.5\textdegree{} (bottom) (see the red zones identified in Figure~\ref{methodomaps}). The segmented images are displayed in the range $[0,1]$ representing the classification value according to which a pixel belongs to the filament class. For each region, the first row shows results of the UNet segmentation, and the second row shows results of the UNet++ segmentation with initial learning-rate values of 10$^{3}$ (left), 10$^{-4}$ to 10$^{-5}$ (right). The UNet++ with a learning rate of 10$^{-2}$ is not presented because of diverging results (see Section \ref{sec:ScoresAnalysis}). The N$_{\rm{H_2}}$ is shown at this position for each region. The regions are $0.45$\textdegree $\times0.45$\textdegree\  wide.}
   \label{res-mods}
    \end{figure*}
In Figure~\ref{fig:PR-curves}, precision-recall curves (P-R curves) are shown for 350.3-353.5\textdegree{} (top left) and 166.1-168.3\textdegree{} (top right). The P-R curve represents precision vs recall for different threshold values. It is used to estimate the optimal threshold that maximizes the dice index (a trade-off between precision and recall). The more the P-R curve tends to the $(1, 1)$ corner, the better the model. In other terms, the larger the area under the curve, the better the model. The objective is to estimate the optimal threshold that returns a trade-off between precision and recall. For clarity sake, all curves are zoomed in from $0-1$ to $0.7-1$ for precision and recall. In all figures, black asterisks refer to the precision-recall values at the optimal threshold; the values of the latter are reported in Tables \ref{tbl:Seg349356}, \ref{tbl:Seg160171}, and \ref{tbl:SegGlobal}. 
Different approaches can be used to compute the optimal threshold, such as minimizing the difference between the precision and recall, or minimizing the Euclidean distance between the P-R curve and the optimal performance, corresponding to a precision-recall of ($1,1$). Here, we computed the optimal threshold as the one that maximized the dice index (trade-off between filament and background recovery). 
In Tables~\ref{tbl:Seg349356} and~\ref{tbl:Seg160171}, we report four samples from these P-R curves corresponding to conservative ($0.8$), medium ($0.5$), relaxed ($0.2$) and optimal (giving the best Dice index) thresholds. When investigating the dense zone of 350.3-353.5\textdegree, we note that, for a given threshold, all the models give results with similar performances (Dice indices $>85\%$), except for UNet[$10^{-5}$] (74.79\% at threshold 0.2). Close optimal threshold are also obtained for all models, where values are situated between $0.35$ and $0.48$. Note that at the conservative threshold 0.8, UNet[$10^{-5}$] and UNet++[$10^{-5}$] result in low recall values compared with the remaining scenarios (72.96\% and 75.14\%, respectively), confirming the results reported with the segmented map where salient filaments are detected with lower values in these two scenarios compared with the remaining ones. However, they tend to be more performing when decreasing classification threshold, especially at threshold 0.2 where they are performing better than the remaining scenarios (the best recall is of 98.1\% with UNet[$10^{-5}$], followed by UNet++[$10^{-5}$] with a recall value of 97.26\%). This can be explained by the low value structures that are better revealed in these two scenarios as noticed before in the segmented map. In Appendix~\ref{tbl:SegGlobal}, precision, recall and Dice index are computed on the fully-segmented Galactic plane, to infer the global segmentation performance. The resulting global scores are inline with the ones obtained with the dense zone where, for a given threshold, all models show close performances, except UNet[$10^{-5}$] and UNet++[$10^{-5}$] slightly less performing. Moreover, the optimal thresholds obtained with the global segmentation are close to thresholds obtained for the dense mosaic, where values range from $0.3$ to $0.44$. This result suggests that either the training step is more driven by high density regions and/or that these regions better represent the global properties observed on the Galactic plane.  \\
\begin{table}[t]
  \caption{Binary Cross Entropy}
  \label{tbl:BCE-test}
  \begin{tabular}{l@{\rule{3em}{0em}}c}
      \hline \textbf{Model} & \textbf{BCE Score} \\
      \hline UNet$[10^{-2}]$ & $0.0085$ \\
      UNet$[10^{-3}]$ & $0.0084$ \\
      UNet$[10^{-4}]$ & $0.0088$ \\
      UNet$[10^{-5}]$ & \textcolor{red}{$0.0161$} \\
      UNet++$[10^{-3}]$ & \textcolor{blue}{$0.0081$} \\
      UNet++$[10^{-4}]$ & $0.0088$ \\
      UNet++$[10^{-5}]$ & $0.0122$ \\
      \hline
  \end{tabular}
  \tablefoot{Binary Cross Entropy (BCE) evaluation on the test set for the schemes reported in Figure~\ref{fig:BCETrainVal}. The lowest (highest) achieved BCE is given in blue (red). BCE values in test are inline with performances in training and validation steps with UNet++[$10^{-3}$] and  UNet[$10^{-5}$] resulting in the best and less performing schemes, respectively.}
\end{table}
When examining the scores of the sparse zone of 166.1-168.3\textdegree{} in Table \ref{tbl:Seg160171}, all the models result in similar performances in precision. However, the recall performance per threshold has a higher contrast, where UNet++[$10^{-5}$] shows lower recall values for thresholds 0.8, 0.5, and 0.2. While the background is well recovered (the lowest precision is of $98.97\%$), lower recall values are obtained compared to the dense zone, where we had to relax the threshold to $0.2$ in order to improve filament recovery and obtain recall values higher than $70\%$. Similarly to the dense zone, close optimal dice indices were obtained for all models, where the difference between the best dice given by UNet++[$10^{-4}$] ($99.38\%$) and the more poorly performing UNet[$10^{-2}$] ($98.8\%$) is less than 0.6\%. Although interesting scores are obtained at the optimal thresholds, it is very important to underline the very low values of these thresholds in all scenarios. Optimal thresholds range from $0.01$ to $< 0.03$, except for UNet[$10^{-5}$] ($0.12$). The obtained values reflect the difficulty of the different networks to reveal structures in this mosaic, where almost $40\%$ to  $60\%$  of the filament pixels are detected with classification values lower than $0.5$ (see the segmented maps in Figure~\ref{res-mods}). Moreover, we clearly note the discrepancy between scores computed on the sparse zone and the global scores reported in Appendices~\ref{tbl:SegGlobal}, which might suggest that the trained models are more successful in revealing filaments in dense than in lower density zones. It is difficult, however, to conclude about the origin of this discrepancy because the number of labeled pixels (in both filament and background) is very different between the sparse and the dense zone. Limited labels in the sparse mosaic also imply that the computed scores with higher uncertainties need to be considered. Nevertheless, visual inspection of the segmented maps in Fig.~\ref{res-mods} results in similar trends as observed for the dense zone, where close performances are noted for all models except for the models with an initial learning rate of $10^{-5}$. These models reveal more details at moderate to low classification values.
\subsection{Analysis of missed structures} \label{sec:MissedStruct}
In addition to pixel-level scores, the structure-level score was also computed in order to evaluate filament recovery in terms of structures. When some pixels of a given filament are missed, it does not automatically imply that the whole structure is missed. In Tables \ref{tbl:Seg349356} and \ref{tbl:Seg160171}, the MS rate is computed at different classification thresholds for dense and sparse zones. As expected, the higher the threshold, the more structures are missed. Overall, we note that at conservative (0.8) and moderate (0.5) thresholds, low MS rates are obtained for dense mosaics compared with the sparse mosaic. In the latter, more strongly contrasting values are obtained across the classification thresholds where MS rates range from 0 (all structures are revealed) to almost 60\% (more than half of the structures are missed; see the MS rates in Table~\ref{tbl:Seg160171}). In a region with a low concentration of filaments, missing (or detecting) a structure would have more impact on the MS rate variation than in a dense region. Similarly to pixel-level scores, the global MS rates reported in Table~\ref{tbl:SegGlobal} are closer to the values obtained with the dense zone than the sparse zone, and this for the same reasons as we invoked for pixel-level scores. 

\begin{figure*}[t]
   %\centering
\includegraphics[width=\linewidth]{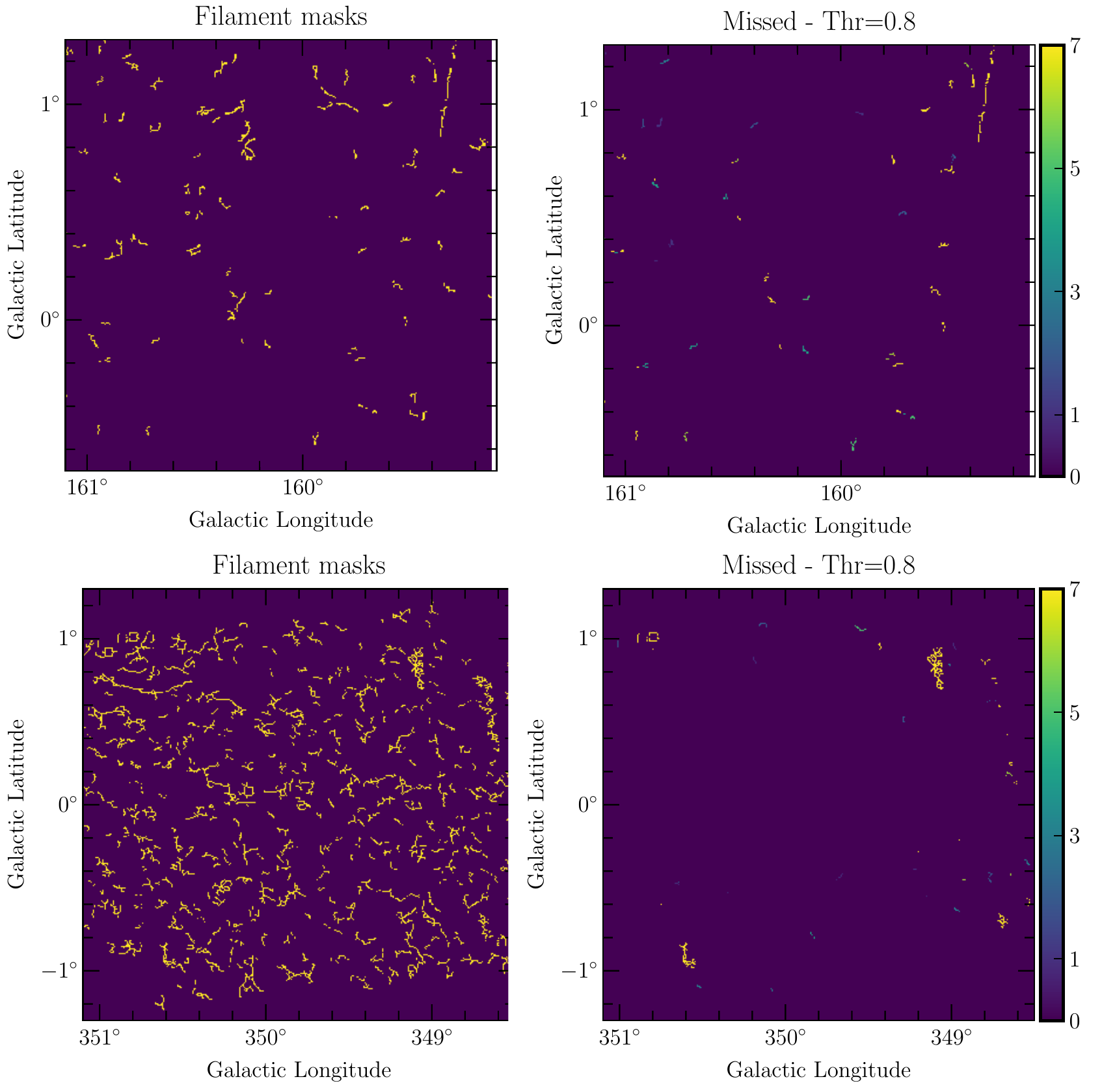}
   \caption{Input filaments missed by the segmentation process on the $l$=160-171\textdegree{} (top) and $l$=349-356\textdegree{} (bottom) regions of the Galactic plane, respectively. We show. the input filament mask (left) and the missed structures at a classification threshold of $0.8$ in a cumulative way (for all the scenarios) (right). The unit (color-coding) for the missed structures maps corresponds to the number of tested scenarios (from 1 to 7) that missed a given structure.}
              \label{missed}%
    \end{figure*}

In order to learn more about the structure of missed filaments, a missed-structure map at a classification threshold of $0.8$ was built. In this map, any structure that was missed by any model is represented. Here, the map intensity encodes the number of models that missed the structure, so that values range from 0 (detected structure) to 7 models (missed by all models). In Figure~\ref{missed}, representative portions from the sparse and dense regions are displayed. We note the prevalence of structures that were missed by all the models (yellow). In fact, $50\%$ of the missed structures in the whole Galactic plane are the same for all the models. After a close visual inspection of the missed structures, two categories are reported. The first category consists of small structures, which is the most prevalent category. These structures either correspond to small isolated filaments and/or to small parts that are missed in larger filaments. The second category corresponds to larger filaments that are misidentified as filaments in the ground truth. For example, structures reported in Figure~\ref{missed} (bottom right) at positions (349\textdegree{}, 1\textdegree) and (350.5\textdegree{}, $-$1\textdegree) are excluded from the filament class. There are also isolated square-shaped structures that are mislabeled as filaments in the ground truth, corresponding to saturated pixels (see Figure~\ref{missed} (bottom right) at position (350.8\textdegree{}, 1\textdegree). Unfortunately, trained models fail to reject these saturation bins when they are entangled within a true large filamentary structure. 
\begin{table*}[h]
\caption{Segmentation scores (350.3-353.5\textdegree{}) }
\label{tbl:Seg349356}
\begin{NiceTabular}{l|*{7}{l}}
\toprule[1pt]\midrule[0.3pt]
\diagbox{Scores (\%)}{Model}   &  UNet [10$^{-2}$] &  UNet [10$^{-3}$] &  UNet [10$^{-4}$] &  UNet [10$^{-5}$] &  UNet++ [10$^{-3}$] &  UNet++ [10$^{-4}$] &  UNet++ [10$^{-5}$] \\
\midrule
Precision at 0.8         &             99.67 &             99.76 &             99.84 &             \bm{\textcolor{blue}{99.89}} &               99.71 &               99.76 &               \textcolor{red}{99.65} \\
Precision at 0.5         &             97.45 &             98.06 &             \textcolor{blue}{98.24} &             97.11 &               97.87 &               97.87 &               \textcolor{red}{95.13} \\
Precision at 0.2         &             88.10 &             89.75 &             88.79 &             60.42 &               \textcolor{blue}{90.12} &               87.59 &               \bm{\textcolor{red}{78.58}} \\
Precision at thr$_{opt}$            &             96.01 &             95.63 &             96.45 &             96.38 &               96.06 &               \textcolor{blue}{96.48} &               \textcolor{red}{94.32} \\ \midrule
Recall at 0.8            &             79.15 &             79.45 &             79.10 &             \bm{\textcolor{red}{72.96}} &               78.34 &               \textcolor{blue}{79.71} &               75.14 \\
Recall at 0.5            &             91.15 &             90.86 &             91.63 &              \textcolor{blue}{91.75} &                \textcolor{red}{90.45} &               91.73 &               90.97 \\
Recall at 0.2            &             96.55 &             96.27 &             96.83 &              \bm{\textcolor{blue}{98.10}} &                \textcolor{red}{96.09} &               96.82 &               97.24 \\
Recall at thr$_{opt}$               &             92.86 &              \textcolor{blue}{93.91} &             93.86 &             92.49 &               93.17 &               93.37 &                \textcolor{red}{91.82} \\ \midrule
Dice at 0.8              &             88.24 &             88.45 &             88.27 &              \textcolor{red}{84.33} &               87.74 &                \textcolor{blue}{88.62} &               85.67 \\
Dice at 0.5              &             94.19 &             94.33 &              \textcolor{blue}{94.82} &             94.35 &               94.02 &               94.70 &                \textcolor{red}{93.00} \\
Dice at 0.2              &             92.13 &             92.89 &             92.64 &              \bm{\textcolor{red}{74.79}} &                \textcolor{blue}{93.01} &               91.98 &               86.92 \\
Dice$_{opt}$                 &             94.41 &             94.76 &              \bm{\textcolor{blue}{95.14}} &             94.40 &               94.59 &               94.90 &                \textcolor{red}{93.05} \\ \midrule
MS at 0.8 &               \textcolor{blue}{4.25} &              4.96 &              4.96 &               \bm{\textcolor{red}{7.79}} &                4.96 &                5.31 &                6.02 \\
MS at 0.5 &               \textcolor{blue}{1.06} &              1.24 &              1.24 &              1.42 &                 \textcolor{red}{1.59} &                1.42 &                 \textcolor{blue}{1.06} \\
MS at 0.2 &               \textcolor{red}{0.35} &               \bm{\textcolor{blue}{0.18}} &               \textcolor{red}{0.35} &               \bm{\textcolor{blue}{0.18}} &                 \textcolor{red}{0.35} &                 \bm{\textcolor{blue}{0.18}} &                 \bm{\textcolor{blue}{0.18}} \\
MS at thr$_{opt}$                   &              0.53 &               \textcolor{blue}{0.35} &              0.53 &               \textcolor{red}{1.24} &                0.53 &                0.53 &                0.88 \\ \midrule\midrule
thr$_{opt}$            &              0.41 &              0.35 &              0.38 &              0.48 &                0.37 &                0.42 &                0.47 \\
\midrule[0.3pt]\bottomrule[1pt]
\end{NiceTabular}
\tablefoot{Segmentation scores evaluated on the the dense zone of 350.3-353.5\textdegree{} segmented by the models reported in Figure \ref{fig:BCETrainVal}. Precision, recall, dice index, and MS rate are evaluated at classification threshold values of $0.8$, $0.5$, $0.2,$ and the optimal threshold. The latter refers to the threshold value optimizing the dice index and estimated using P-R curves (see Figure \ref{fig:PR-a}). Blue (red) refers to the best (least performing) model in each row. The bold scores correspond to the absolute best (if in blue) or lowest (if in red) performance per score. }
\end{table*}

\begin{table*}[t]
\centering
\caption{Segmentation scores (166.1-168.3\textdegree)} \label{tbl:Seg160171}
\begin{NiceTabular}{l|*{7}{l}}
\toprule[1pt]\midrule[0.3pt]
\diagbox{Scores (\%)}{Model}     &  UNet [10$^{-2}$] &  UNet [10$^{-3}$] &  UNet [10$^{-4}$] &  UNet [10$^{-5}$] &  UNet++ [10$^{-3}$] &  UNet++ [10$^{-4}$] &  UNet++ [10$^{-5}$] \\
\midrule
Precision at 0.8         &  \bm{\textcolor{blue}{100.00}} & \bm{\textcolor{blue}{100.00}} &  \bm{\textcolor{blue}{100.00}} &            \bm{\textcolor{blue}{100.00}} &  \bm{\textcolor{blue}{100.00}} &  \bm{\textcolor{blue}{100.00}} &   \bm{\textcolor{blue}{100.00}} \\
Precision at 0.5         &  \bm{\textcolor{blue}{100.00}} &  \bm{\textcolor{blue}{100.00}} &  \bm{\textcolor{blue}{100.00}} &   \bm{\textcolor{blue}{100.00}} &  \bm{\textcolor{blue}{100.00}} &    \bm{\textcolor{blue}{100.00}} &    \bm{\textcolor{blue}{100.00}} \\
Precision at 0.2         &  \bm{\textcolor{blue}{100.00}} &    \bm{\textcolor{blue}{100.00}} &   \bm{\textcolor{blue}{100.00}} &    \bm{\textcolor{blue}{100.00}} &  \bm{\textcolor{blue}{100.00}} &  \bm{\textcolor{blue}{100.00}} &   \bm{\textcolor{blue}{100.00}} \\
Precision at thr$_{opt}$            &             99.04 &    \bm{\textcolor{red}{98.97}} &             99.65 & \textcolor{blue}{99.66} &               99.25 &               99.32 &               99.24 \\ \midrule
Recall at 0.8            &             25.91 &             28.38 &             25.22 &             25.29 &               \textcolor{blue}{33.45} &               30.64 &               \bm{\textcolor{red}{20.84}} \\
Recall at 0.5            &             49.01 &             58.40 &             49.01 &             54.90 &               \textcolor{blue}{63.33} &               60.59 &               \textcolor{red}{41.74} \\
Recall at 0.2            &             77.11 &             87.32 &             81.49 &             \textcolor{blue}{94.59} &               87.18 &               87.73 &               \textcolor{red}{72.86} \\
Recall at thr$_{opt}$               &             \textcolor{red}{98.56} &             98.83 &             98.90 &             99.04 &               99.31 &               \bm{\textcolor{blue}{99.45}} &               99.04 \\ \midrule
Dice at 0.8              &             41.15 &             44.21 &             40.28 &             40.37 &               \textcolor{blue}{50.13} &               46.90 &               \bm{\textcolor{red}{34.49}} \\
Dice at 0.5              &             65.78 &             73.73 &             65.78 &             70.88 &               \textcolor{blue}{77.55} &               75.46 &               \textcolor{red}{58.90} \\
Dice at 0.2              &             87.07 &             93.23 &             89.80 &             \textcolor{blue}{97.22} &               93.15 &               93.46 &               \textcolor{red}{84.30} \\
Dice$_{opt}$                 &             \textcolor{red}{98.80} &             98.90 &             99.28 &             99.35 &               99.28 &               \bm{\textcolor{blue}{99.38}} &               99.14 \\ \midrule
MS at 0.8 &         \bm{\textcolor{red}{61.02}} &             45.76 &             57.63 &             50.85 &               \textcolor{blue}{32.20} &               37.29 &               44.07 \\
MS at 0.5 &             \textcolor{red}{16.95} &              \textcolor{blue}{6.78} &             10.17 &             15.25 &                \textcolor{blue}{6.78} &                \textcolor{blue}{6.78} &               11.86 \\
MS at 0.2 &              \textcolor{red}{1.69} &     \bm{\textcolor{blue}{0.00}} &              \textcolor{red}{1.69} &              \bm{\textcolor{blue}{0.00}} &                \bm{\textcolor{blue}{0.00}} &                \textcolor{red}{1.69} &                \textcolor{red}{1.69} \\
MS at thr$_{opt}$                   &   \bm{\textcolor{blue}{0.00}} &             \bm{\textcolor{blue}{0.00}} &              \bm{\textcolor{blue}{0.00}} &              \bm{\textcolor{blue}{0.00}} &                \bm{\textcolor{blue}{0.00}} &                \bm{\textcolor{blue}{0.00}} &                \bm{\textcolor{blue}{0.00}} \\ \midrule\midrule
thr$_{opt}$            &              0.03 &              0.03 &              0.02 &              0.12 &                0.01 &                0.02 &                0.02 \\
\midrule[0.3pt]\bottomrule[1pt]
\end{NiceTabular}
\tablefoot{Same as in Table~\ref{tbl:Seg349356}, but for the sparse zone of 166.1-168.3\textdegree}
\end{table*}

%!!!!!!!!!!!!!!!!!!!!!!!!!!!!!!!!!!!!!!!!!!!!!!!!!!!!!!!!!!!!!!!!!!!!!!!!!!!!!!!!!!!!!!!!!!!!!!!!!!!!!!!!!!!!
\subsection{New possible filaments revealed by deep learning} \label{sec:AstroResults}
\begin{figure*}[h]
\includegraphics[width=\linewidth]{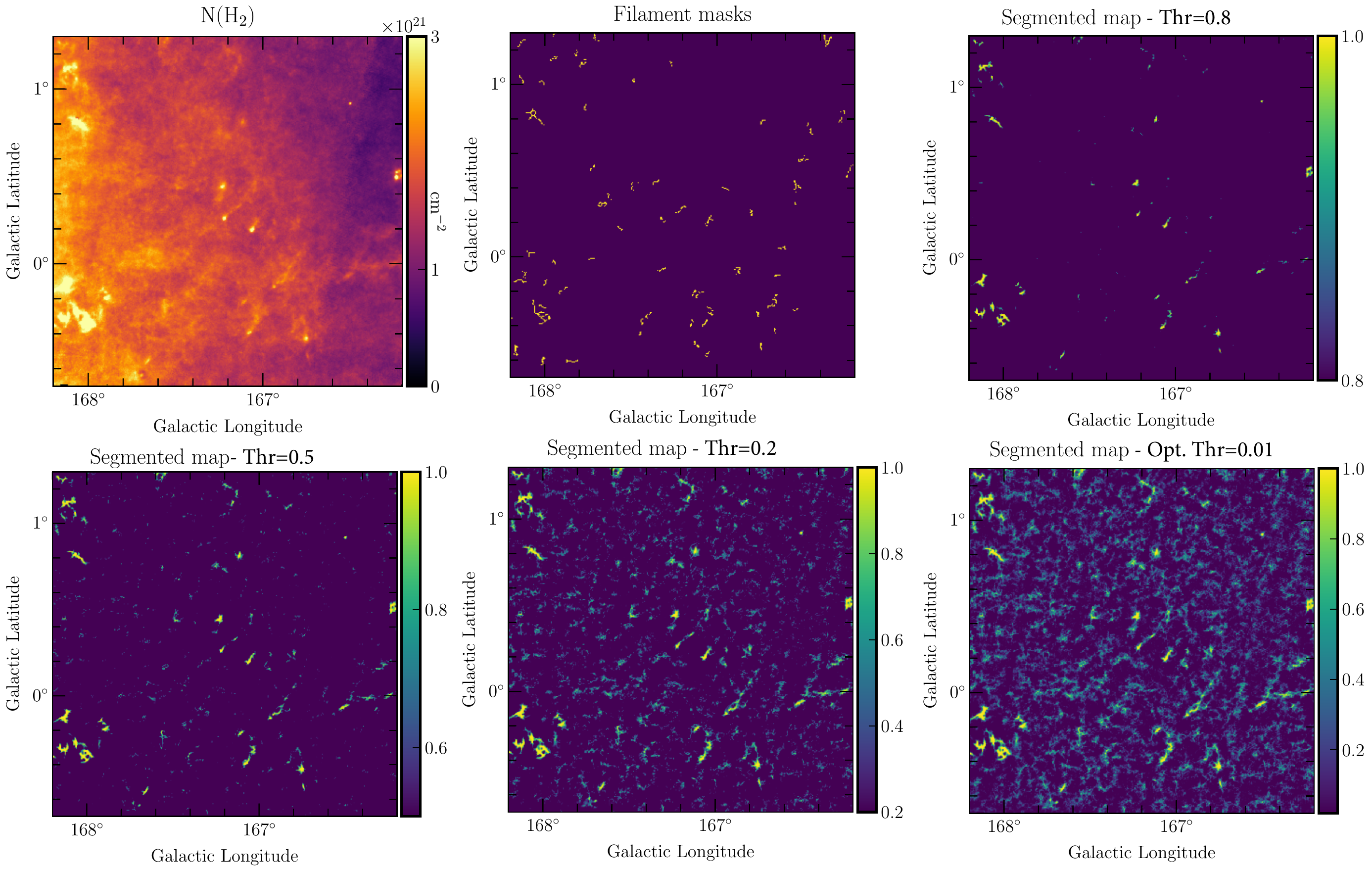}
\includegraphics[width=\linewidth]{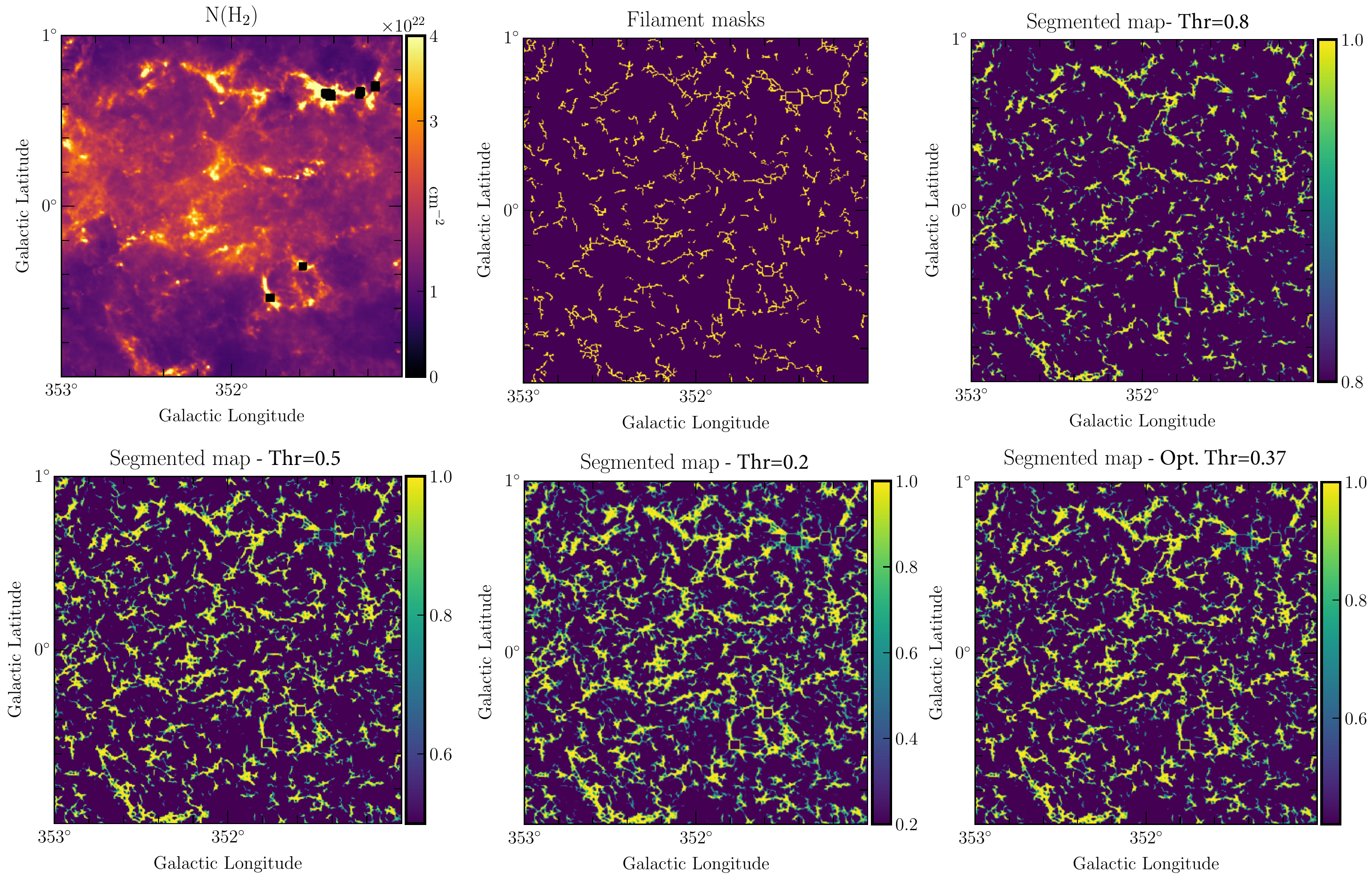}
 \caption{Zoom-in on the evolution of segmentation results as a function of the classification threshold showing the filamentary structures estimated by UNet++ 10$^{-3}$ in 160-171\textdegree{} (six top images) and 349-356\textdegree{} (six bottom images) regions of the Galactic plane. The original H$_2$ column density image (top left in each group), the ground-truth input filament mask, and the corresponding segmented image at different thresholds (0.8, 0.5, 0.2, and the optimal threshold) are shown from top left to bottom right. The regions are $2$\textdegree $\times2$\textdegree\ wide.}
\label{threshold}%
\end{figure*}

Based on the performance analysis in Section~\ref{sec:ScoresAnalysis}, two groups of models can be derived based on the initial learning rate: 1) models with an initial learning rate of $10^{-4},10^{-3},\text{and }10^{-2}$ , and 2) models with an initial learning rate of $10^{-5}$. In the following, we present the segmentation results that we illustrate for the best model, UNet++[$10^{-3}$], on the two selected submosaics. \\
Figure~\ref{threshold} presents the evolution of the segmented map for the two selected regions as a function of the segmented map threshold. 

\begin{figure*}[t]
     \begin{subfigure}{0.5\textwidth}
         \includegraphics[width=\linewidth]{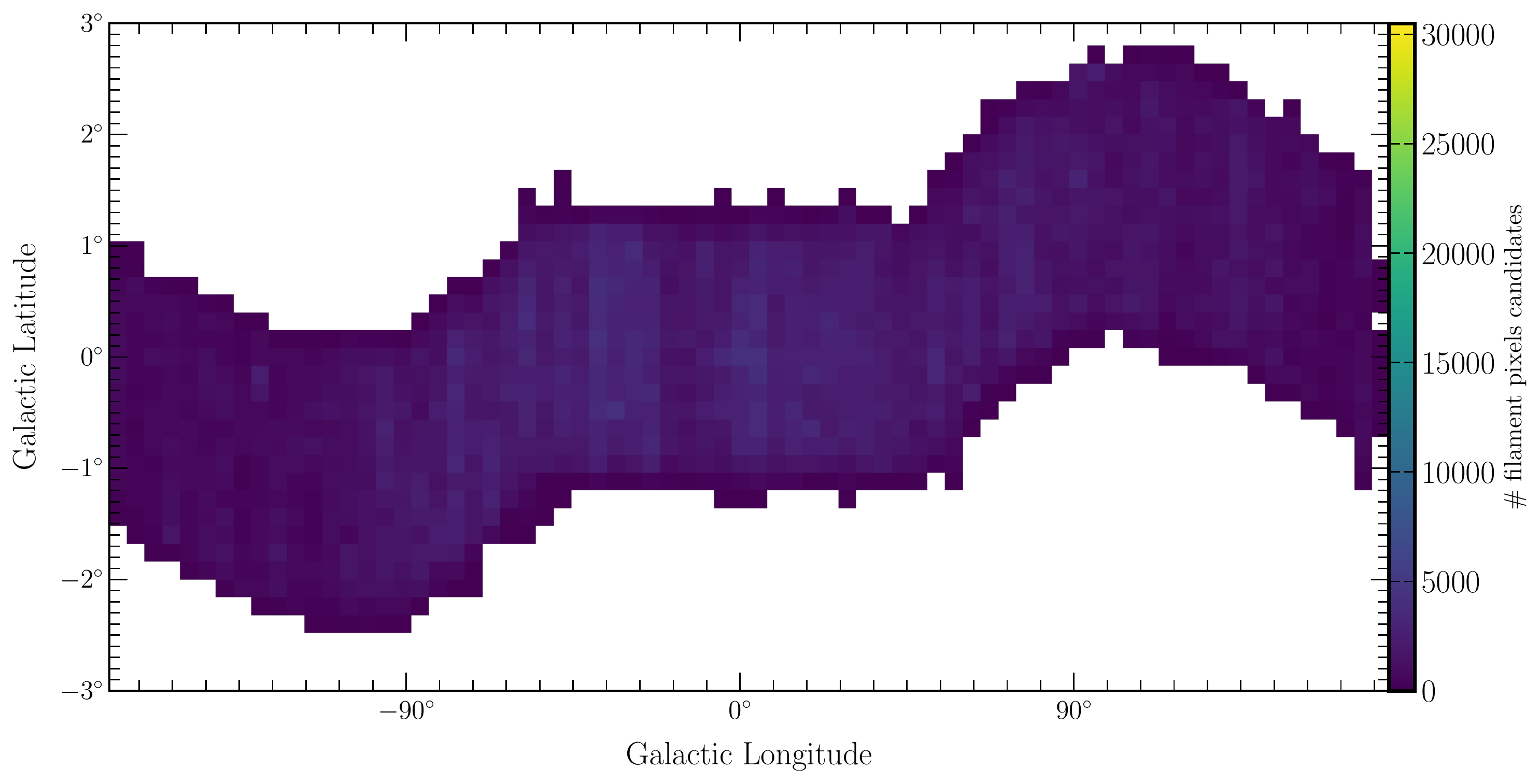}
         \caption{Ground truth}
         \label{fig:Groundtruth}
     \end{subfigure}
     \begin{subfigure}{0.5\textwidth}
         \includegraphics[width=\linewidth]{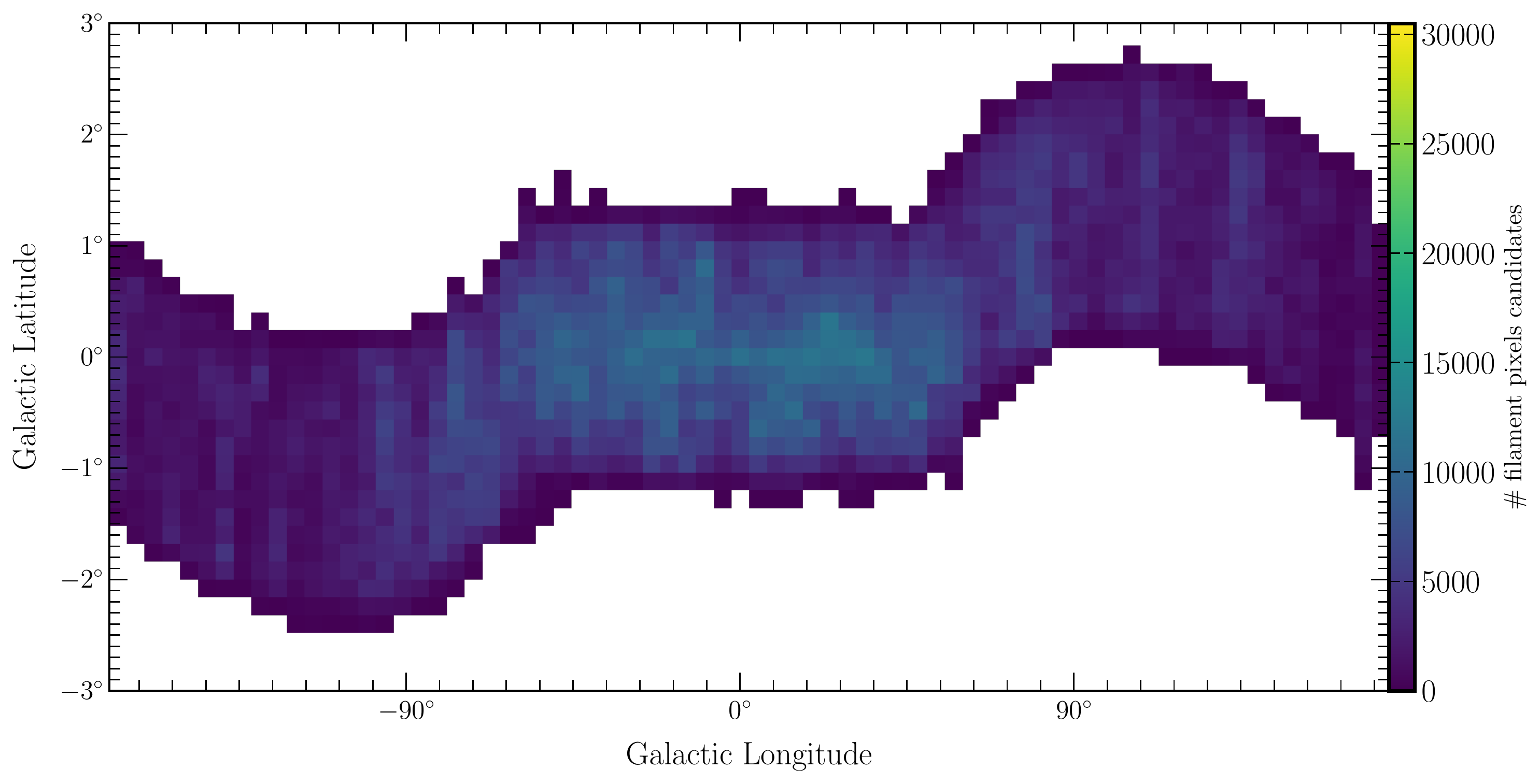}
         \caption{Segmentation at classification threshold of 0.8}
         \label{fig:Seg0p8}
     \end{subfigure} 
     \newline
     \begin{subfigure}{0.5\textwidth}
         \includegraphics[width=\linewidth]{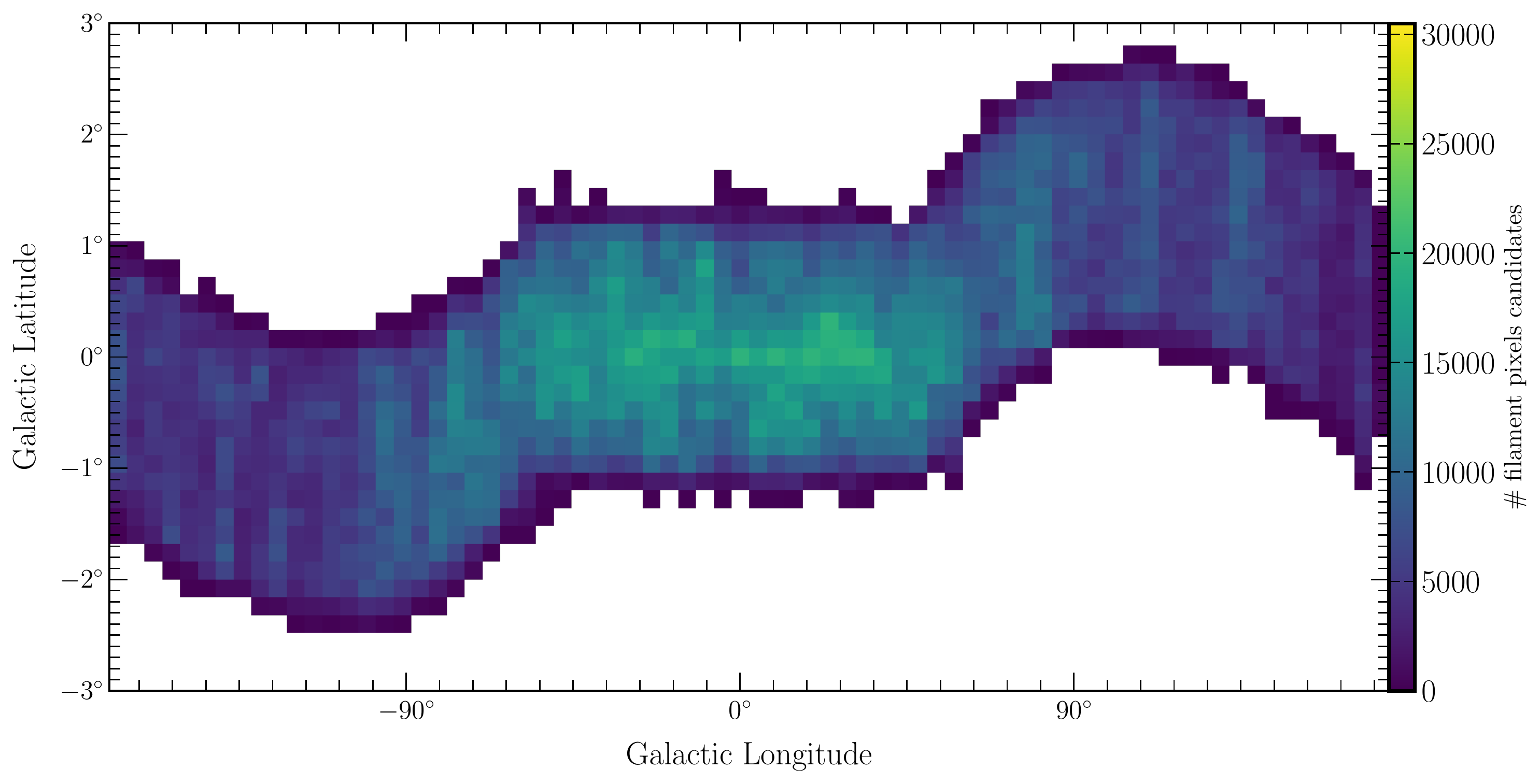}
         \caption{Segmentation at classification threshold of 0.5}
         \label{fig:Seg0p5}
     \end{subfigure}
     \begin{subfigure}{0.5\textwidth}
         \includegraphics[width=\linewidth]{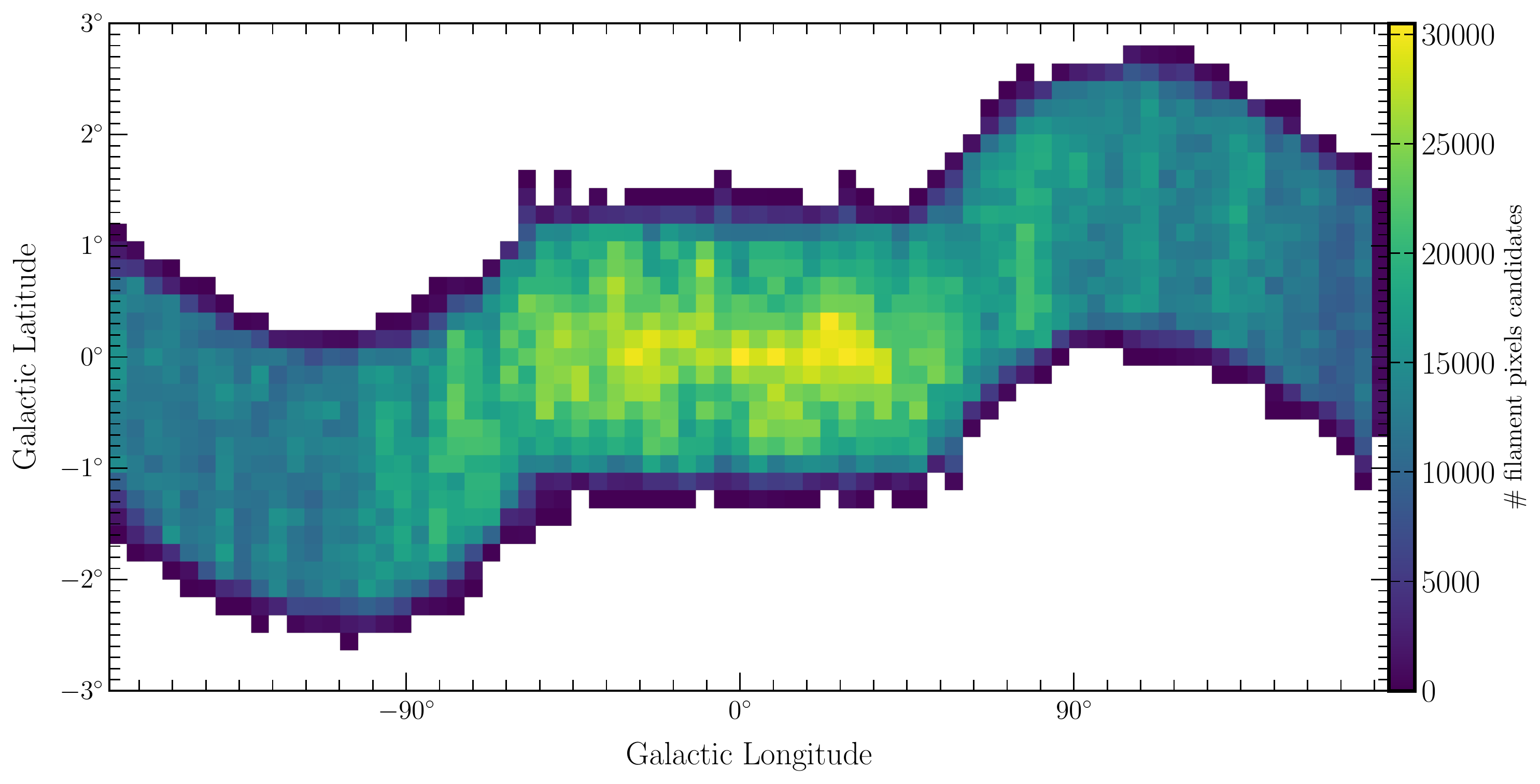}
         \caption{Segmentation at classification threshold of 0.2}
         \label{fig:Seg0p2}
     \end{subfigure}
     \newline
     \begin{subfigure}{0.5\textwidth}
         \includegraphics[width=\linewidth]{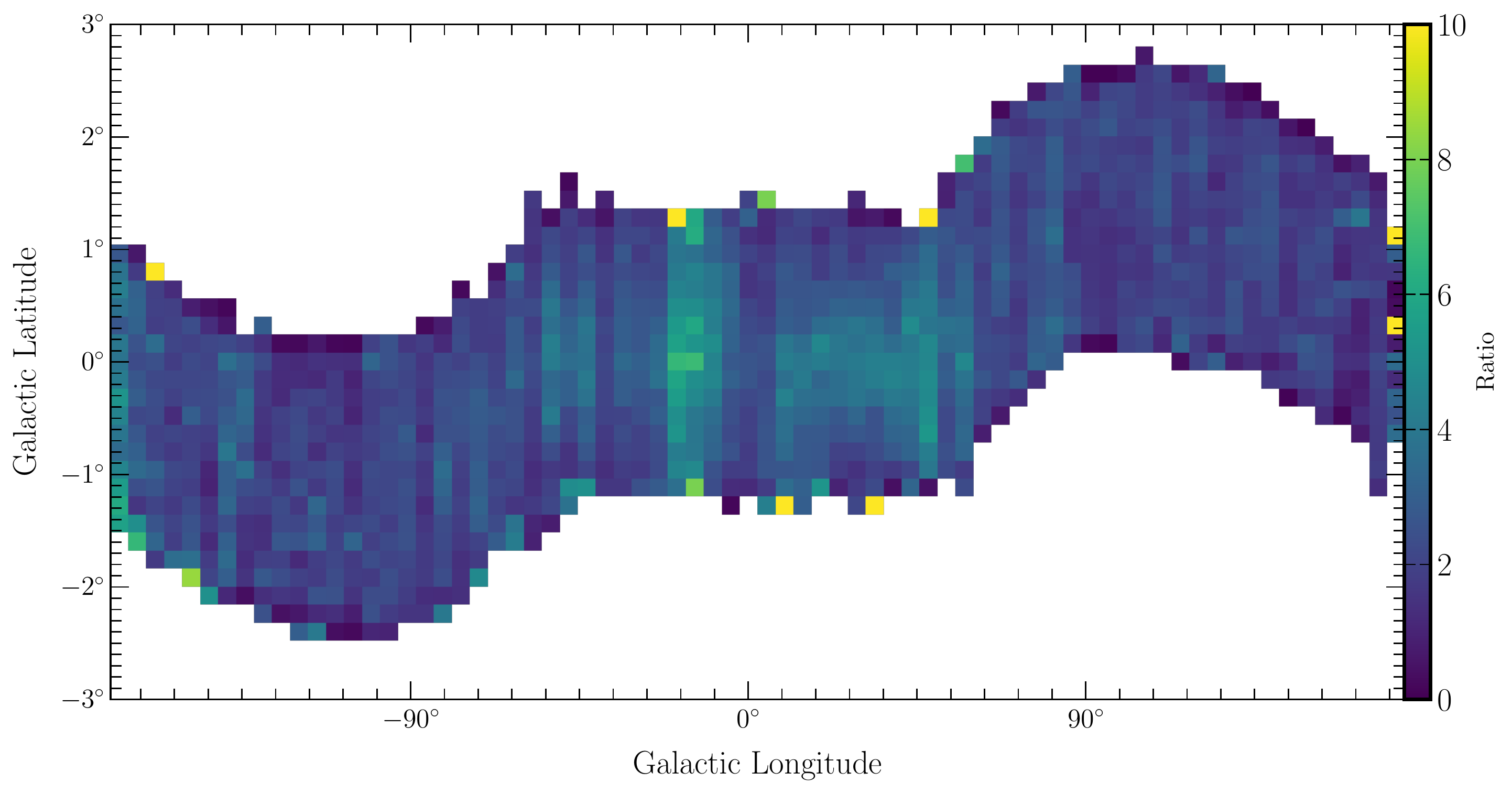}
         \caption{Ratio of \ref{fig:Seg0p8} to \ref{fig:Groundtruth} }
         \label{fig:Ratio}
     \end{subfigure}
        \caption{Number density distribution of candidate filament pixels across the entire Galactic plane, estimated in bins of 4.8\textdegree{} $\times$ 0.16\textdegree{} and comparing the ground truth of Fig.~\ref{fig:Groundtruth} with the segmentation results of the model UNet++ 10$^{-3}$ at classification thresholds of 0.8 in Fig.~\ref{fig:Seg0p8}, 0.5 in Fig.~\ref{fig:Seg0p5}, and 0.2 in Fig.~\ref{fig:Seg0p2}. In Fig.~\ref{fig:Ratio}, we display the ratio of candidate filament pixels in the segmentation at a classification threshold of 0.8 to the ground truth.}
        \label{fig:PDF_pixels}
\end{figure*}

\begin{figure*}[h]
\includegraphics[width=\linewidth]{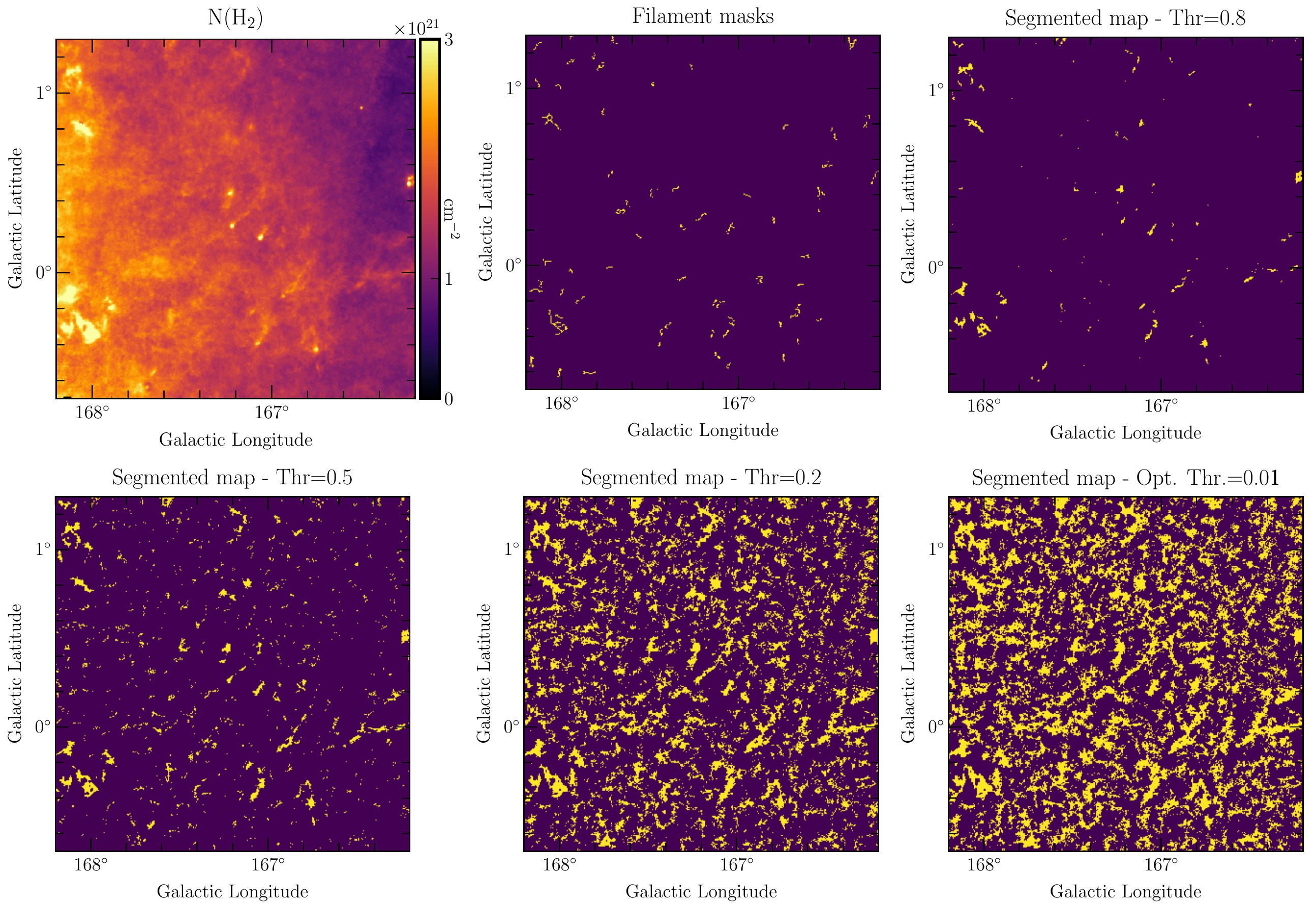}
\includegraphics[width=\linewidth]{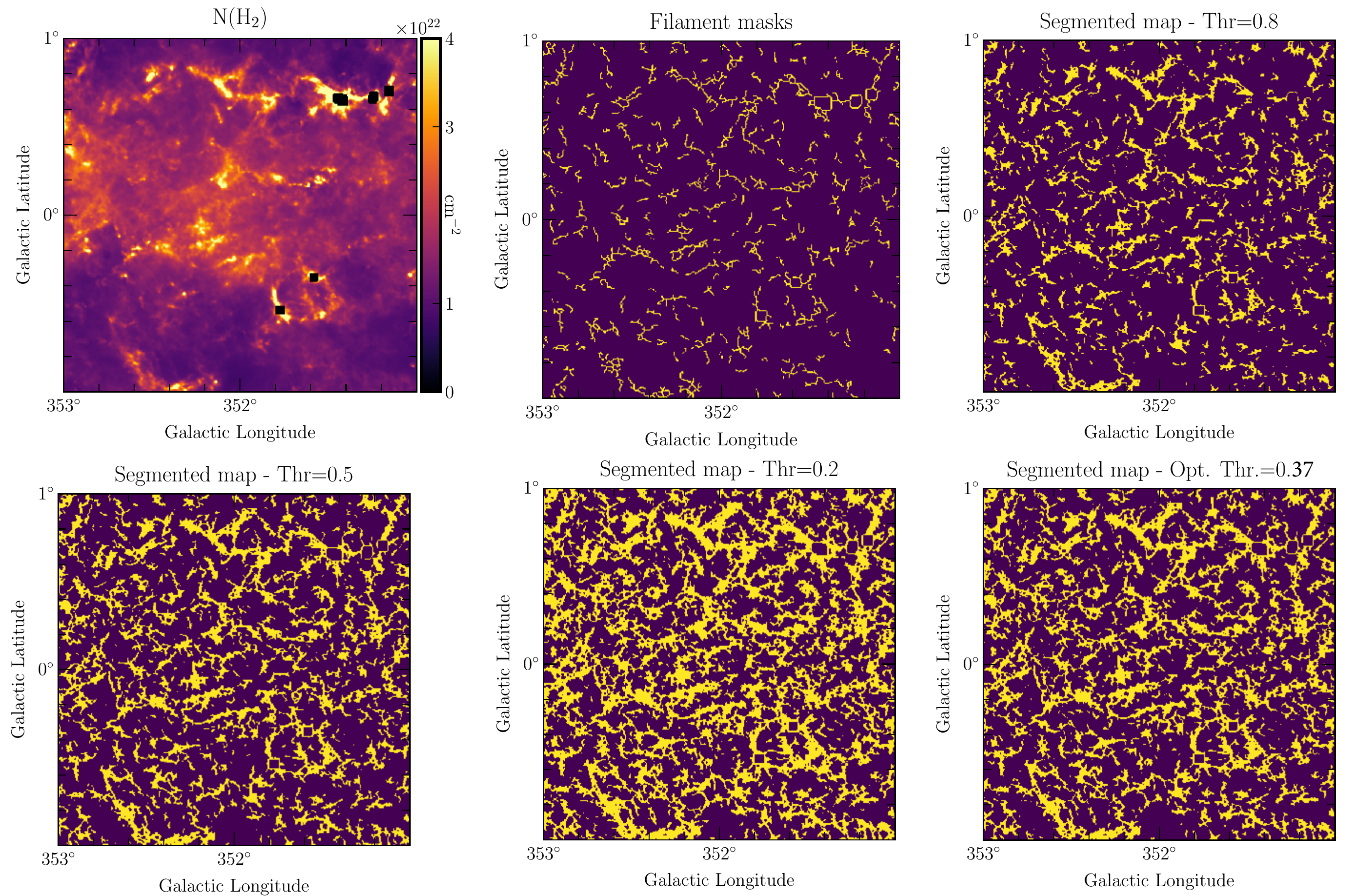}
 \caption{Zoom-in on the evolution of segmentation results generated by UNet++ 10$^{-3}$ model in mosaics 160-171\textdegree{} (six top images) and 349-356\textdegree{} (six bottom images). Here, the estimated binary filament masks are displayed at classification thresholds 0.8, 0.5, 0.2, and the optimal threshold. The original H$_2$ column density image (image with the color bar) and the true input filament mask are also displayed for comparison. The regions are $2$\textdegree $\times2$\textdegree\  wide. }
\label{threshold-bin}%
\end{figure*}

\begin{figure}[htb]
   \centering
\includegraphics[width=\linewidth]{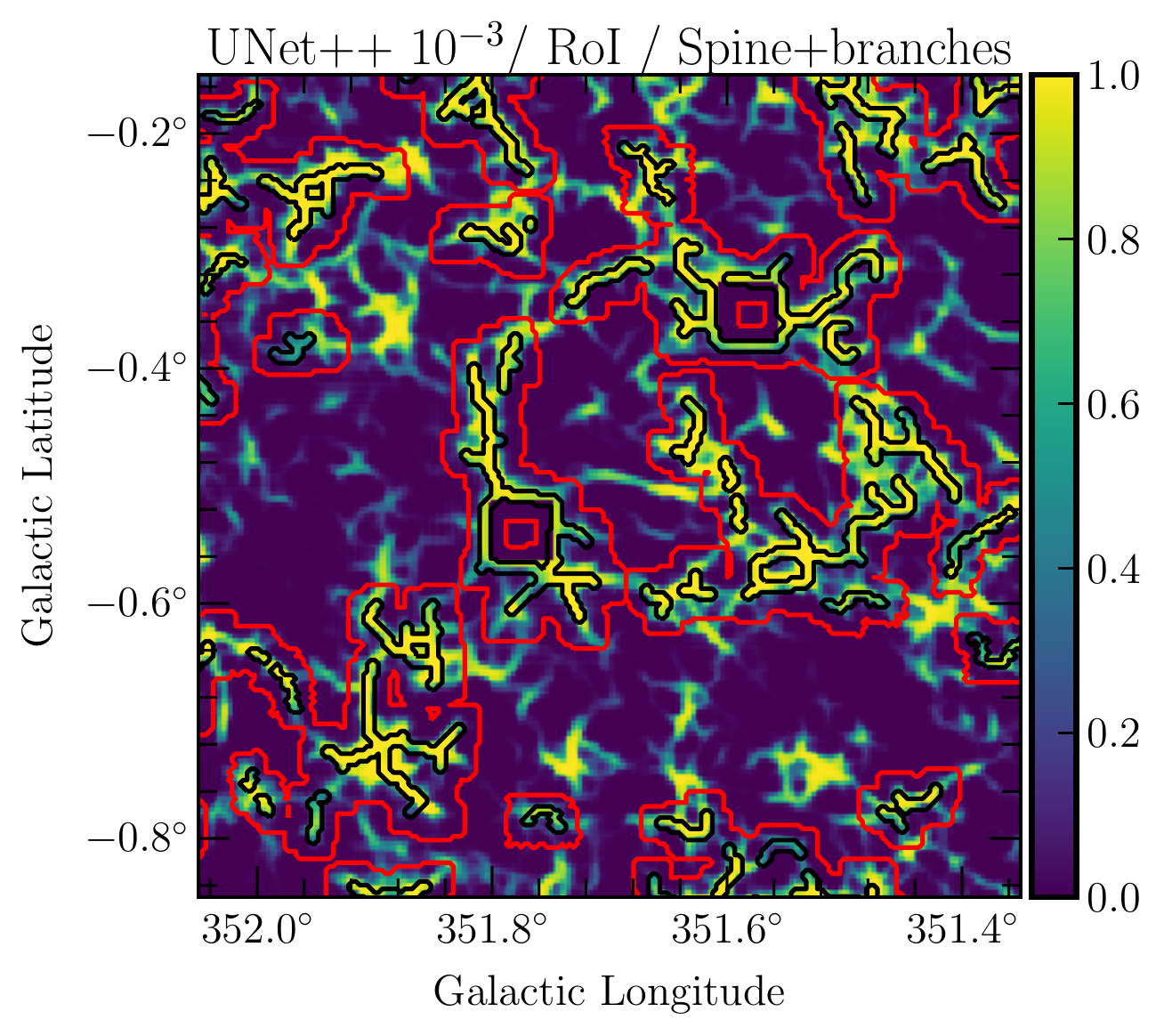}
   \caption{Comparison of the segmented map result using the UNet++ 10$^{-3}$ with the input spine plus branches (black contours) and the RoI (red contours).  }
              \label{Unet-RoI-spines}%
    \end{figure}

In both regions, more pixels are classified as filaments by the training and segmentation processes than in the input structures (input filament mask used as the ground truth in the supervised training). The ratio (new filament pixels to input filament pixels) varies between 2 to 7, depending on the threshold value (from 0.8 to 0.3, the optimal threshold). The same conclusions are drawn for the whole Galactic plane. In Figure~\ref{fig:PDF_pixels}, the distribution of candidate filament pixels across the entire Galactic plane, estimated in bins of 4.8\textdegree{} $\times$ 0.16\textdegree{}, is displayed for the ground truth and the segmentation results at different classification thresholds. Even at a conservative classification threshold of 0.8, more pixels are labeled as filaments than in the ground truth used in the learning step. As expected, the more we decrease the classification threshold, the more pixels are labeled as filament.
A close visual inspection of the segmented images indicates that structures observed at thresholds lower than the optimal value are also seen on the original column density image, but were not previously detected due to their low contrast with respect to the surrounding background emission.
The squared structures shown in Figure~\ref{threshold} (bottom) are saturated pixels corresponding to bright sources located in filaments. These structures also appear in the ground truth, and we therefore retrieve them when applying our model. Because we lack information about the column density in these saturated pixels, we left them as squares in the segmented maps. \\
Figure~\ref{threshold-bin} presents the same result as Figure~\ref{threshold}, but for the binarized version of the segmented map. 
The filamentary structures identified at a given threshold are now represented as 1 when the associated pixel belongs to the filament class 0 instead. This representation allows a more direct comparison with the input filament mask (spine plus branches; see Fig.~\ref{fil}). However, the classification value itself (that indicates whether a pixel belongs to the filament class) no longer appears in this representation. As for results presented in Figure~\ref{threshold}, the threshold decrease has two effects: i) more pixels are identified as belonging to the filament class (new structures are detected, in particular,  those with a faint-to-low contrast that are barely visible in the original N$_{\rm{H_2}}$ image), and ii) a given structure becomes thicker. This last effect can be identified as a lowering of the corresponding column density threshold, where the highest threshold identifies the densest part of the filament. It is interesting to note that the optimal threshold (as well as lower threshold values) in both regions identifies the filamentary structures as observed in the original column density map, down to their external envelope emission, before reaching the background emission. This result is important because it will allow a precise study of the filament-background relation. The widening of the filamentary structures can also be seen as the definition of the RoI given by \citet{schisano2020} (see also Figure~\ref{fil}, bottom right).  \citet{schisano2020} defined the RoI as the objects that define filamentary candidates in their catalog. This point is important because it implies that the comparison of our segmented map results with the RoI would lower the factor we derived that represents the number of pixels classified as filament as the RoI are always thicker than the spine plus branches we used as input in this work to train the network to learn what a filament is. In this work, we infer the filament mask as a ground truth from the filament spine plus branches, as shown on Figure~\ref{fil} (bottom left). However, the RoI defines the filament in order to delineate its spread on the column density map, and this corresponds to the observed widening of the structures with the lowering of the threshold. The comparison of the different structures is illustrated in Figure~\ref{Unet-RoI-spines}. 
Because we used spine plus branches as input to define a filament here, we kept this input structure as a reference to compare with the result of the segmentation process.         

\begin{figure*}[h]
  % \centering
\includegraphics[width=\linewidth]{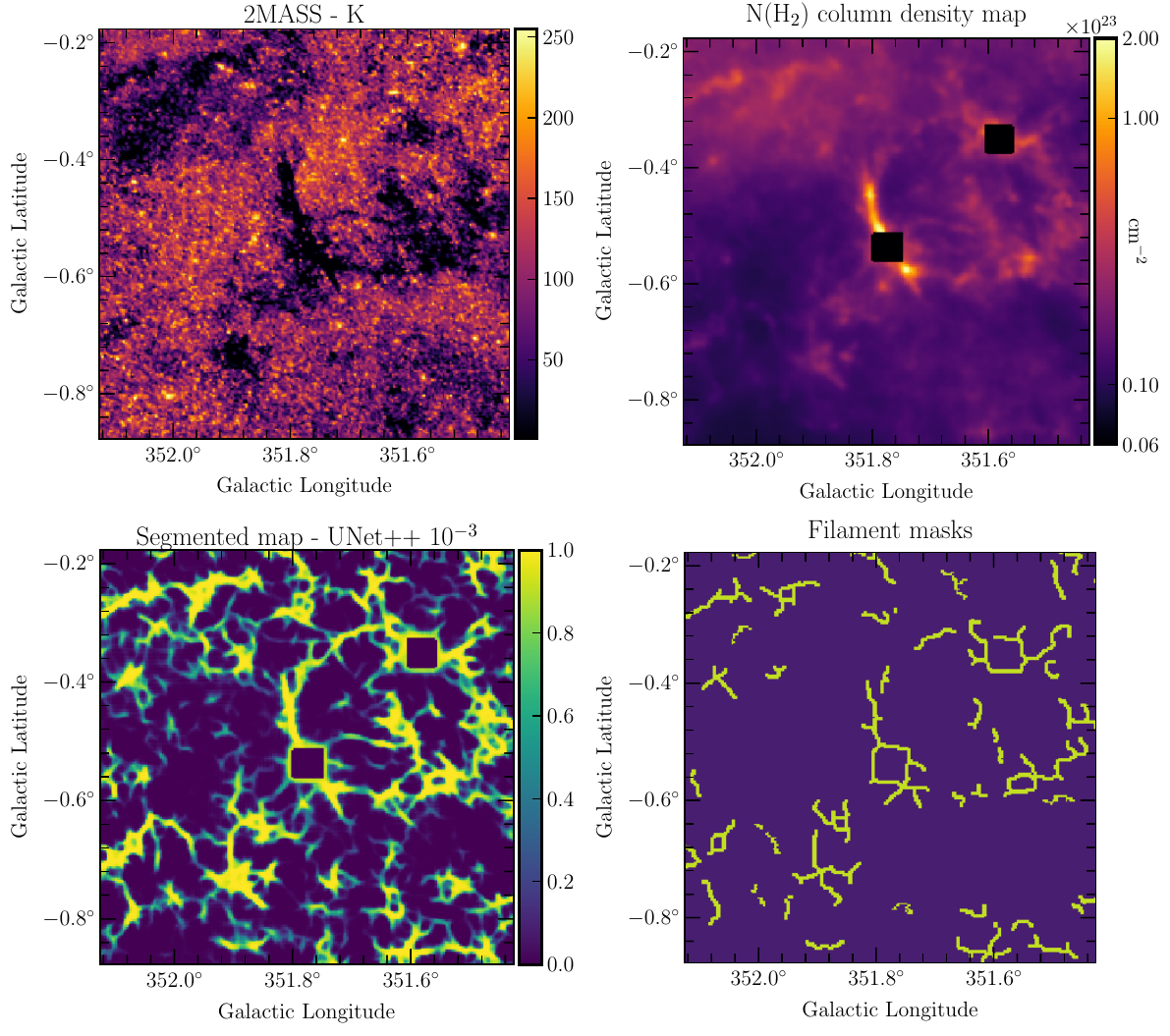}
 \caption{Segmented map obtained for star-forming region G351.776-0.527 (bottom left) compared with the 2MASS $K$-band image (top left), where filaments are observed in absorption, and the column density map (top right), where filaments are observed in emission. The filamentary structures visible in the image segmented by UNet++[10${-3}$] (bottom left) and displayed in the [0,1] range are both visible in the 2MASS $K$ and N$_{\rm{{H_2}}}$ column density images, ensuring their nature. The G351.776-0.527 source is located at the center of the images, connected to a filamentary network (hub). The corresponding input filament mask is shown (bottom right).}
              \label{G351}
   \end{figure*}

A key point in this work is to ascertain the nature of the filamentary structures we reveal. Filaments are made of gas and dust. The filaments detected by \citet{schisano2020} are traced in dust emission maps. Dust grains emit over a wide range of wavelengths, and they act as absorbers in the optical, near-, and mid-infrared parts of the spectrum. Because of their dusty composition, filaments are well visible as absorbing features at shorter wavelengths (optical, near-, and mid-infrared) in the Galactic plane, and these data can be used to ascertain the nature of the structures returned by the training and segmentation processes. Only filaments that are visible in absorption on a strong emission background can be detected at short wavelengths. (Sub)millimeter emission of cold dust also reveals filaments \citep{mat18,leu19} and can be used to ascertain the nature of the structures without encountering the extinction problem. This empirical (data-based) validation of the results is a first step in the analysis. In Figure~\ref{G351}, we illustrate the interest of this multiwavelength analysis to ascertain the nature of new detected filaments on the Galactic region G351.776-0.527. This region hosts a high-mass star-forming region analyzed by \citet{leu19}.
Region G351.776-0.527 is located at the center of Figure~\ref{G351} and appears as a hub with filaments converging toward the saturated central point located at $l$=351.77\textdegree, $b=-0.538$\textdegree. A filamentary structure observed in the segmented map right of the bright central source that is not visible in the input filament mask is seen in the 2MASS $K$-band image, confirming its nature. The segmented image also suggests that the region might be located at the edge of a bubble. A bright ionized region, G351.46-00.44, is located nearby and could explain the high level of turbulence and the high-mass star formation observed in this zone \citep{lee12}. The large-scale view of this region, revealed by the segmented map with suggested multiple filament connections of the central source with the surrounding medium, has to be confirmed with high-sensitivity observations of dust emission that could be complemented by spectroscopic data of dense gas molecular tracers, keeping in mind that, as pointed by \citet{Hac22}, filaments identified with \textit{Herschel} data (i.e., using dust continuum emission) might be a different family of objects than those detected in molecular line tracers. 
\section{Discussion and future prospects}\label{dis}
The purpose of this work was to study the potential of supervised deep learning as a new way to detect filaments in images of the Galactic interstellar medium. At this stage, the filamentary structures are \textit{\textup{revealed,}} but the filaments themselves are not \textit{\textup{extracted}}, with a measurement of their physical properties, from the segmented images. While the first task requires semantic segmentation, the second task consists of instance segmentation \citep{GU2022}. In the first, the task is limited to attributing a class to each pixel, which is done in this paper, whereas in the latter, existing filaments are in addition enumerated to allow a global statistical study. In this paper, we used UNet-based networks which are the most modern methods in semantic segmentation. The analyzed performance in Section~\ref{res} proves the efficiency of these networks not only in revealing structures already existing in the initial catalog, but also in adding new structures that have not been detected before and that are confirmed through a detection at shorter and/or longer wavelengths, namely, at near- and mid-infrared and/or (sub)millimeter wavelengths, respectively. In astrophysics, several independent estimators can be used to ascertain the true nature of the detected filamentary structures, such as expert knowledge or a knowledge based on a large statistical definition, such as the one used in citizen projects, of particular interest for machine learning \citep{Chr22}. Results of numerical simulations and/or data obtained at other wavelengths can also be used. On the multiwavelength data side, for example, filaments are clearly visible at other wavelengths in the Galactic plane because they are composed of dust, and these data can be used to ascertain the nature of the structures returned by the segmentation process, as shown in Figure~\ref{G351}. \\
The UNet-based networks are supervised deep-learning algorithm. In spite of the incomplete ground truth, these networks produced a good estimate of filamentary structures. It is important to build a more enriched ground truth, however, to solve more complex tasks such as instance segmentation. This might be possible by combining several existing catalogs of filaments obtained on the Galactic plane, for example, by combining the Hi-GAL catalog \cite{schisano2020} with \textit{getSF} extractions made on several regions of the plane \citep{men21}. Another possibility is to use filament segmentation by UNets as a prestep and then consider the produced filaments mask as the ground truth for the instance segmentation. \\
Another crucial step in filament segmentation using deep-learning algorithms is data normalization. Filament detection depends not only on the intrinsic column density of the structure, but also on the column density and the structure of the background. By using local (per patch) normalization, the filament contrast relative to the neighboring background is enhanced. As illustrated in Figure~\ref{fig:normalization}, a classical local normalization method was used in this work in order to enhance low-contrast filaments that are in turn well integrated in the training process, allowing them to be representative. Recently, more sophisticated multiscale normalization has been used in the Hi-GAL image processing to highlight the faintest structures observed on the Galactic plane \citep{licausi2016}. These normalized data are very interesting as input for deep-learning networks. Unfortunately, no ground truth exists for these normalized images so far, which does not allow their use for the moment. \\
An important result of the segmentation for astrophysical purposes is to determine the classification threshold (intensity of the segmentation map) that allows for an optimal detection of filaments. While the optimal thresholds reported in Table \ref{tbl:SegGlobal} allowed a good recovery of existing and new filaments, filaments are still missed at these classification thresholds, some because they were misclassified in the ground truth (see Sect.~\ref{sec:MissedStruct}). \\
Another key point is to consider a region-specific optimal threshold rather than a unique global one. According to results reported in Tables~\ref{tbl:Seg349356}, \ref{tbl:Seg160171}, and \ref{tbl:SegGlobal}, the optimal threshold of a given model is affected by the column density of the studied zone. In this work, the optimal thresholds in Tables \ref{tbl:Seg349356} and \ref{tbl:Seg160171} were inferred in narrow zones, with a low number of labeled pixels for the sparse zone. To obtain a robust estimate of a region-specific optimal threshold, a split of the Galactic plane divided into large homogeneous zones in terms of filament concentration is envisioned. The optimal thresholds can then be inferred on these slices. 

From the computational point of view, we trained the different networks on a NVIDIA RTX 2080Ti. Table~\ref{tbl:Training-Time} gives the training time for the different scenarios for 100 epochs. As UNet++ is a larger network than UNet, it takes slightly more time to train: while UNet requires about $2.2$h, UNet++ asks for about $2.8$h. As the patch data set is small (around 210~MB), these times are not impacted by the loading of the data. When the training step is completed, the neural networks are usually faster on CPU than on GPU~\cite{goodfellow2016deep} because the transfer from CPU memory to GPU memory takes time. The segmentation of the mosaics was therefore built on an Intel CPU machine (i7-10610U). It took about $4$h per mosaic with an overlap of 30 pixels and 32$\times$32 patches. The total training and segmentation time is therefore estimated to be $6.5$h per mosaic.     

\begin{table}[t]
\centering
  \caption{Training time}
  \label{tbl:Training-Time}
  \begin{tabular}{l@{\rule{3em}{0em}}c}
      \hline \textbf{Model} & \textbf{Training time [hour]} \\
      \hline UNet$[10^{-2}]$ & \textcolor{blue}{$2.08$} \\
      UNet$[10^{-3}]$ & $2.23$ \\
      UNet$[10^{-4}]$ & $2.19$ \\
      UNet$[10^{-5}]$ & $2.17$ \\
      UNet++$[10^{-3}]$ & \textcolor{red}{$3.22$} \\
      UNet++$[10^{-4}]$ & $2.79$ \\
      UNet++$[10^{-5}]$ & $2.73$ \\
      \hline
  \end{tabular}
  \tablefoot{Training time in hours for the schemes reported in Figure \ref{fig:BCETrainVal}. The shortest (longest) training time, in blue (red), is achieved by UNet[$10^{-2}$] (UNet++[$10^{-3}$]).}
\end{table}
From the method point of view, future works will include an improvement of the segmentation process by using dedicated windows to build the patches as in~\cite{pielawski2020introducing}. These tools may dramatically lower the computational burden. We also investigate alternative ways of building a larger set of patches while keeping good statistical properties. The method used in this work  guarantees a good preservation of the statistical distribution, but leads to a small number of patches.

Although this method has some limitations, that is, the limited quantity of patches for the training and the  incomplete ground truth, and although it reveals filament structures instead of extracting them, the net increase (a factor between 2 and 7 on the whole segmented map) of the number of pixels that belongs to the filament class and the robust detection of intrinsically faint and/or low contrast ones offers important perspectives. We currently explore the implementation of an augmented ground-truth data set using results of numerical simulations on Galactic filaments. The extraction and separation of filament pixels observed in the 2D segmented map is also ongoing, and we add 3D spectroscopic data.

\section{Conclusions}\label{conc}
We explored whether deep-learning networks, UNet-5 and UNet++, can be used to segment images of the whole Galactic plane in order to reveal filamentary structures. 
\begin{itemize}
\item Using molecular hydrogen column density image of the Galactic plane obtained as part of the Hi-GAL survey and filaments previously extracted by \citet{schisano2020}, we trained two different UNet-5 based networks with six different scenarios based on a different initial learning rate. 
\item We showed the results and estimated the performances of the different scenarios that we presented for two representative mosaics of the Galactic plane selected for their low and high column density density and filament content. 
\item We determined the best models for these mosaics based on machine-learning metrics. We focused the training estimates on the recovery of input structures (filaments and background) and defined for each mosaic and for the whole plane an optimal classification threshold that ensured the best recovery of input structures.  
\item We show that depending on the model and the selected threshold, new pixels classified as filament candidates increase by a factor between 2 to 7 (compared to the input spine+branches structures used as ground truth).\ This suggests that this new method has the potential of revealong filamentary structures that may not be extracted by non-ML-based algorithms. 
\end{itemize}
We point out the high potential of the produced database for future studies of filaments (statistical analysis or follow-ups). We will use the results of the numerical simulations to enrich the ground truth and assess the uncertainties on segmented maps. The astrophysical analysis of the produced database is ongoing and will be published in a separate paper.  

\begin{acknowledgements}
      This project has received financial support from the CNRS through the MITI interdisciplinary programs (Astroinformatics 2018, 2019). This project is part of the ongoing project BigSF funded by the Aix Marseille Université A*Midex Foundation (SB post-doctoral funding). AZ thanks the support of the Institut Universitaire de France. 
      We thank the anonymous referee for their constructive comments that helped to improve the quality of the paper. This research has made use of "Aladin Sky Atlas" developed at CDS, Strasbourg Observatory, France.
\end{acknowledgements}
\bibliographystyle{aa} % style aa.bst
\bibliography{Fil-ML} 
\begin{appendix} 
\section{Different input data maps} \label{appendixA}
%\subsection{Merging individual maps}
All the input maps used in the training process (the column density map, the filament mask, the background mask, and the missing data mask maps; see Section~\ref{datasetbuilding}) were originally produced from the individual Hi-GAL mosaics extending over 10\textdegree\, of longitude (see \citet{schisano2020}). To facilitate the use of the four input maps during the training process, we merged the individual maps using the Python module \textit{reproject}\footnote{\url{https://reproject.readthedocs.io/en/stable/index.html}}. The patches were then cut as shown in Figure~\ref{fig:patches_overlap}. The four input maps used in the supervised training process are presented in Figure~\ref{methodomaps}. These maps are described below. 

\subsection*{N$_{\rm{H_2}}$ map}
The column density maps N$_{\rm{H_2}}$ were obtained as part of the \textit{Herschel} observations of the Galactic plane, Hi-GAL. The column density N$_{\rm{H_2}}$ maps were computed from the photometrically calibrated Hi-GAL mosaics following the
approach described in \citet{eli13}. The \textit{Herschel} data were convolved to the 500\,$\mu$m resolution (~36$\arcsec$) and rebinned on that map grid. Then a pixel-by-pixel fitting
with a single-temperature greybody function was performed, as described in \citet{eli13,schisano2020}. We directly used the data derived and presented in \citet{schisano2020}.  

\subsection*{Missing-data map}
\begin{figure}[h]
     \centering
     \begin{subfigure}[b]{0.4\linewidth}
         %\centering
         \includegraphics[width=\linewidth]{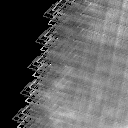}
         \caption{}
         \label{fig:missing_boundary}
     \end{subfigure}
     \hfill
     \begin{subfigure}[b]{0.4\linewidth}
         %\centering
         \includegraphics[width=\linewidth]{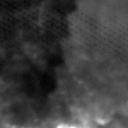}
         \caption{}
         \label{fig:missing_noise}
     \end{subfigure} \\
     \begin{subfigure}[b]{0.4\linewidth}
         %\centering
         \includegraphics[width=\linewidth]{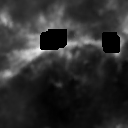}
         \caption{}
         \label{fig:missing_saturation}
     \end{subfigure}
     \hfill
     \begin{subfigure}[b]{0.4\linewidth}
         %\centering
         \includegraphics[width=\linewidth]{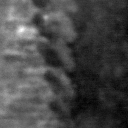}
         \caption{}
         \label{fig:missing_structure}
     \end{subfigure}
        \caption{Four sources of missing data. (a) Example of a complex boundary due to satellite scanning. (b) Example of a noisy pattern inside the column density map (in log scale) due to the satellite scan. (c) Saturated pixels. (d) Example of structured artifacts built by the mapping process.}
        \label{fig:missing_data}
\end{figure}

The N$_{\rm{H_2}}$ map presents local degraded zones (noise, saturation, and overlap issues) that we wished to exclude from the training process. Examples of these degraded zones are shown in Figure~\ref{fig:missing_data}. 
In Figure~\ref{fig:missing_data} we show the four sources of missing data. First, the boundaries of the mosaics show a grid-like pattern, as shown in Figure~\ref{fig:missing_boundary}. These features are due to the scanning pattern followed by the \textit{Herschel} satellite while performing the observations of an Hi-GAL tile. The grid pattern corresponds to the slewing phase of the scanning pattern, when the satellite inverted the scan direction. Second, the Hi-GAL mosaics were built using the Unimap mapmaking \citep{tra11} following a strategy to avoid large-scale intensity gradients over the image. 
We refer to Appendix A of \citet{schisano2020} for a detailed description of how the mosaics were built. In short, UNIMAP mapmaker was run to simultaneously process the data of two adjacent Hi-GAL tiles, creating what the authors call a texel, spanning 4\textdegree $\times$ 2\textdegree. The UNIMAP processing is able to deal with the slewing region along the side covered by observation of both Hi-GAL tiles. This produces an image of better quality with respect to the simple mosaicking of the two tiles. Multiple texels are combined together to produce the overall mosaic that is 10\textdegree{} wide in longitude. While it is possible to directly process a larger portion of the Galactic plane with UNIMAP, this approach introduces large gradients in intensity over the entire mosaic. The UNIMAP mapmaker produces maps with an average intensity level equal to zero. Therefore, the texels require to be calibrated in flux \citep{ber10}. 
In some cases, there are discrepancies in the calibration level of adjacent texels that introduce sharp variations in intensity in the overlapping region between texels. An example is show in Figure~\ref{fig:missing_noise} and Figure~\ref{fig:missing_structure}. These variations are not physical and were masked out from the learning process.

The final full map of missing data was built by combining all the types of possible missing data described above. They were then removed from the training. 

\subsection*{Filament-mask map}
\citet{schisano2020} published a catalog of filaments detected in the Galactic plane from N$_{\rm{H_2}}$ images obtained with Hi-GAL photometric data. The position of spines and branches is available through binary masks in which pixels belonging to these structures are tagged for the 32059 filaments published in the catalog (see Figure~\ref{fil} and their figure 3). We used this information as an input mask to define our ground truth for the filaments. The ground truth was defined over the outputs of the Hi-GAL filament catalog, and it depends on the completeness of that catalog. This implies that the ground truth is not absolutely fully defined because it will miss the information from any feature that may not have been detected by \citet{schisano2020}. 

\subsection*{Background map}
On all pixels, the column density mosaic contains emission from both filaments and background \citep{schisano2020}. We examined each 10\textdegree{} mosaic to define a background level as the lowest level of emission observed on the mosaic that does not overlap any filament branches that were detected and used as ground truth. This means that our definition of the background emission sets the background class on a very low number of pixels, but ensures that these pixels do not contain filaments. We work on a more precise definition of the background to allow more pixels to be labeled in this class. We are primarily interested in detecting filamentary structures here. Moreover, the \textit{\textup{local}} normalization applied to all patches before the training process (see Figure~\ref{fig:normalization}) tends to limit the impact of the background on the detection of filaments. This reduces the need for a precise definition of the background further.  

\section{Supplementary quantitative scores} \label{appendixB}
\begin{table*}[h]
\centering
  \caption{Dice index}
  \label{tbl:Dice-test}
  \begin{tabular}{l@{\rule{3em}{0em}}l@{\rule{3em}{0em}}l@{\rule{3em}{0em}}l@{\rule{3em}{0em}}l}
      \hline \textbf{Model} & \textbf{Dice index [$0.2$]} & \textbf{Dice index [$0.4$]} & \textbf{Dice index [$0.6$]} & \textbf{Dice index [$0.8$]}\\
      \hline UNet$[10^{-2}]$ & $93.19$ & $94.34$ & $94.07$ & $92.06$ \\
      UNet$[10^{-3}]$ & $93.13$ & $94.37$ & $94.16$ & \textcolor{blue}{$92.43$} \\
      UNet$[10^{-4}]$ & $92.75$ & $94.09$ & $93.79$ & $91.9$ \\
      UNet$[10^{-5}]$ & \bm{\textcolor{red}{$82.4$}} & \textcolor{red}{$89.6$} & $91.3$ & \textcolor{red}{$87.66$} \\
      UNet++$[10^{-3}]$ & \textcolor{blue}{$93.46$} & \bm{\textcolor{blue}{$94.61$}} & \textcolor{blue}{$94.28$} & $92.36$ \\
      UNet++$[10^{-4}]$ & $93.04$ & $94.07$ & $93.6$ & $91.57$ \\
      UNet++$[10^{-5}]$ & $90.24$ & $91.88$ & \textcolor{red}{$91.02$} & $87.67$ \\
      \hline
  \end{tabular}
  \tablefoot{Dice index evaluation on the test set for the schemes reported in Figure \ref{fig:BCETrainVal} at classification threshold values of $0.8$, $0.6$, $0.4,$ and $0.2$. Blue (red) refers to the best (less performing) scheme in each column. The bold scores correspond to the absolute best (if in blue) or lowest (if in red) dice index. The closer the dice index to $1$, the better. Dice index values in test were aligned with performances in training and validation steps as close performances are obtained for schemes with initial learning-rate values in $[10^{-4},10^{-3},10^{-2}]$ and lower performance are noted for schemes with a learning-rate value of $10^{-5}$. Moreover, UNet++[$10^{-3}$] and  UNet[$10^{-5}$] result in the best ($94.61\%$ at $0.4$) and least performing ($82.4\%$ at $0.2$) schemes, respectively.}
\end{table*}
\begin{table*}[h]
\centering
  \caption{Segmentation scores}
  \label{tbl:SegGlobal}
\begin{NiceTabular}{l|*{7}{l}}
\toprule[1pt]\midrule[0.3pt]
\diagbox{Scores (\%)}{Model}     &  UNet [10$^{-2}$] &  UNet [10$^{-3}$] &  UNet [10$^{-4}$] &  UNet [10$^{-5}$] &  UNet++ [10$^{-3}$] &  UNet++ [10$^{-4}$] &  UNet++ [10$^{-5}$] \\
\midrule
Precision at 0.8         &             \textcolor{red}{99.71} &             99.74 &             \bm{\textcolor{blue}{99.75}} &             99.73 &               99.74 &               99.74 &               99.72 \\
Precision at 0.5         &             98.29 &             98.45 &             \textcolor{blue}{98.54} &             \textcolor{red}{97.07} &               98.47 &  98.42 &               97.35 \\
Precision at 0.2         &             91.43 &             92.08 &             91.75 &      \bm{\textcolor{red}{64.85}} &  \textcolor{blue}{92.45} &               91.17 &               86.16 \\
Precision at thr$_{opt}$            &             95.65 &             95.85 &             95.88 &             94.95 &               95.75 &               \textcolor{blue}{95.94} &               \textcolor{red}{93.61} \\ \midrule
Recall at 0.8            &             \textcolor{blue}{72.38} &             71.34 &             70.50 &            \bm{\textcolor{red}{65.14}} &               70.87 &               71.74 &               65.54 \\
Recall at 0.5            &             \textcolor{blue}{90.13} &             89.90 &             89.13 &             89.25 &               89.78 &               89.78 &               \textcolor{red}{85.59} \\
Recall at 0.2            &             97.49 &             97.64 &             97.29 &             \bm{\textcolor{blue}{98.72}} &               97.61 &               97.50 &               \textcolor{red}{95.81} \\
Recall at thr$_{opt}$               &             95.26 &             95.50 &             94.96 &             92.24 &               \textcolor{blue}{95.67} &               94.97 &               \textcolor{red}{91.56} \\ \midrule
Dice at 0.8              &             \textcolor{blue}{83.88} &             83.19 &             82.62 &             \textcolor{red}{78.81} &               82.87 &               83.46 &               79.10 \\
Dice at 0.5              &             \textcolor{blue}{94.05} &             94.00 &             93.62 &             93.01 &               93.95 &               93.92 &               \textcolor{red}{91.11} \\
Dice at 0.2              &             94.41 &             94.82 &             94.49 &             \bm{\textcolor{red}{78.31}} &               \textcolor{blue}{95.01} &               94.27 &               90.76 \\
Dice$_{opt}$                 &             95.46 &             95.68 &             95.42 &             93.58 &              \bm{\textcolor{blue}{ 95.71}} &               95.45 &               \textcolor{red}{92.57} \\ \midrule
MS at 0.8 &             \textcolor{blue}{14.36}&             14.81 &             16.51 &             17.58 &               15.28 &               15.10 &               \bm{\textcolor{red}{17.69}} \\
MS at 0.5 &              4.61 &              4.40 &              5.25 &              \textcolor{blue}{4.07} &                4.62 &                4.55 &                \textcolor{red}{5.83} \\
MS at 0.2 &              0.83 &              0.77 &              0.81 &              \bm{\textcolor{blue}{0.28}} &                0.85 &                0.82 &                \textcolor{red}{1.10} \\
MS at thr$_{opt}$                   &              1.95 &              1.69 &              1.88 &              2.72 &                \textcolor{blue}{1.68} &                1.97 &                \textcolor{red}{2.85} \\ \midrule\midrule
thr$_{opt}$            &              0.32 &              0.31 &              0.31 &              0.44 &                0.30 &                0.33 &                0.34 \\
\midrule[0.3pt]\bottomrule[1pt]
\end{NiceTabular}
\tablefoot{Segmentation scores evaluated on the Galactic plane segmented by the models reported in Figure \ref{fig:BCETrainVal}. Precision, recall, dice index, and MS rate are evaluated at classification threshold values of $0.8$, $0.5$, $0.2,$ and the optimal threshold. The latter refers to the threshold value optimizing the dice index and estimated using P-R curves (see Figure \ref{fig:PR-c}). Blue (red) refers to the best (least performing) scheme in each row. The bold scores correspond to the absolute best (if in blue) or lowest (if in red) performance per score. Overall, the segmentation scores obtained on the fully segmented Galactic plane consolidate the results obtained with the zone of 350.3-353.5\textdegree{} in Table~\ref{tbl:Seg349356}}
\end{table*}
\end{appendix}
\end{document}